\documentclass[10pt,journal]{IEEEtran}
\usepackage{cite}
\ifCLASSINFOpdf
\else
  \usepackage[dvips]{graphicx}
  \DeclareGraphicsExtensions{.eps}
\fi
\usepackage[cmex10]{amsmath}
\usepackage{amsfonts,amssymb,amsthm,mdwmath,bbm}
\newtheorem{lemma}{Lemma}

\usepackage{array}

\ifCLASSOPTIONcompsoc
  \usepackage[caption=false,font=normalsize,labelfont=sf,textfont=sf]{subfig}
\else
  \usepackage[caption=false,font=footnotesize]{subfig}
\fi

\usepackage{stfloats}
\usepackage{url}
\usepackage{color}
\usepackage{acronym}
\begin{document}
\acrodef{PPP}[PPP]{Poisson Point Process}
\acrodef{NPPP}[NPPP]{Non-homogeneous PPP}
\acrodef{PGFL}[PGFL]{Probability Generating Functional}
\acrodef{CDF}[CDF]{Cumulative Distribution Function}
\acrodef{PDF}[PDF]{Probability Distribution Function}
\acrodef{PMF}[PMF]{Probability Mass Function}
\acrodef{PCF}[PCF]{Pair Correlation Function}
\acrodef{RV}[RV]{Random Variable}
\acrodef{SIR}[SIR]{Signal-to-Interference Ratio}
\acrodef{i.i.d.}[i.i.d.]{independent and identically distributed}
\acrodef{w.r.t.}[w.r.t.]{with respect to}
\acrodef{MAC}[MAC]{Medium Access Control}
\acrodef{V2V}[V2V]{Vehicle-to-Vehicle}
\acrodef{1D}[1D]{one-dimensional}
\acrodef{VANET}[VANET]{Vehicular ad hoc network}
\acrodef{LT}[LT]{Laplace Transform}
\acrodef{CoV}[CoV]{Coefficient-of-Variation}

\title{The Meta Distribution of the SIR in Linear Motorway VANETs}

\author{Konstantinos Koufos and Carl P. Dettmann % <-this % stops a space
\thanks{K.~Koufos and C.P.~Dettmann are with the School of Mathematics, University of Bristol, BS8 1TW, Bristol, UK. \{K.Koufos, Carl.Dettmann\}@bristol.ac.uk} \protect \\ 
\thanks{This work was supported by the EPSRC grant number 
EP/N002458/1 for the project Spatially Embedded Networks. All underlying data are provided in full within this paper.}}

\maketitle

\begin{abstract}
The meta distribution of the signal-to-interference-ratio (SIR) is an important performance indicator for wireless networks because, for ergodic point processes, it describes the fraction of scheduled links that achieve certain reliability, conditionally on the point process. In this paper, we calculate the moments of the meta distribution in vehicular ad hoc networks (VANETs) along high-speed motorways. Due to the high speeds, the drivers maintain large safety distances, and the Poisson point process (PPP) becomes a poor deployment model. Because of that, we model the distribution of inter-vehicle distance equal to the sum of a constant hardcore distance and an exponentially distributed random variable. We design a novel \emph{discretization model} for the locations of vehicles which can be used to approximate well the meta distribution of the SIR due to the hardcore process. We validate the model against synthetic motorway traces. On the other hand, the PPP overestimates significantly the coefficient-of-variation of the meta distribution due to the hardcore process, and its predictions fail. In addition, we show that the calculation of the meta distribution becomes especially meaningful in the upper tail of the SIR distribution. 
\end{abstract}

\begin{IEEEkeywords}
Headway distance models, probability generating functional, reduced Palm measure, synthetic mobility traces.
\end{IEEEkeywords}
\section{Introduction}
The long-term vision of having vehicles communicating with each other for improving traffic flow, enabling automated driving, etc. is not far from reality~\cite{Filippi2016}. The first standardization actions started in 1999, once the Federal Communication Commission in U.S. allocated 75 MHz of spectrum in the 5.9 GHz band for dedicated short-range communication~\cite{FCC2003}. In 2008, the European Commission set aside 30 MHz for cooperative intelligent transport systems. Since 2010, the technology amendment IEEE 802.11p has been the basis for world-wide PHY/MAC layer standards supporting \ac{V2V} communication in the 5.9 GHz band. In addition, the \ac{V2V} communication will be secured under the umbrella of cellular LTE networks~\cite{Chen2017}. 

The performance of \acp{VANET} has been extensively studied during the past three decades using measurements and simulations, see for instance~\cite{Fiore2014,Fiore2015} and the references therein. Unfortunately, both methods lack scalability. Recently, analytical tools, like spatial point processes~\cite{Veres2002}, have been employed to gain quick insights into the system performance~\cite{Renzo2015}. The classical analysis of wireless networks using stochastic geometry assumes a spatial model for the network elements, and averages the performance indicator (mostly outage probability) over all network states~\cite{Andrews2011}. The average does not represent well the reliability of each individual link, when the standard deviation (of the indicator) is comparable to the mean. Because of that, the meta distribution of the \ac{SIR} has been recently proposed to assess the distribution of the outage probability, conditioned on the realization of the point process~\cite{Ganti2010}. Thus far, the meta distribution of bipolar, cellular and heterogeneous wireless networks has been investigated~\cite{Haenggi2016, Haenggi2018}. 

The spatial distribution of vehicles requires a model for the road infrastructure and another for the locations of vehicles conditionally on the roads. The Manhattan Poisson line process (with horizontal and vertical layout of streets) and the Poisson line process (for streets with random orientation) are popular in urban vehicular communication studies. For analytical tractability, they are coupled with \ac{1D} \ac{PPP} for the locations of vehicles along the streets. The resulting point process is commonly referred to as a Cox process in the plane. The study in~\cite{Baccelli2015} shows that the distribution of interference level is discontinuous at the intersections, the study in~\cite{Dhillon2018} brings up the trade-off between the intensities of streets and vehicles in the coverage probability (or probability of successful reception) of the typical receiver, and the study in~\cite{Baccelli2018} enhances the model of~\cite{Dhillon2018} assuming both vehicular and macro-base stations. Simpler models for the road network, e.g., two orthogonal streets in~\cite{Steinmetz2015} and a grid of roads in~\cite{Haenggi2017}, highlight the fact that the coverage probability of the typical receiver becomes lower near intersections, because there, the generated interference from both horizontal and vertical streets is significant.

A \ac{1D} setup should suffice for the modeling of a motorway, and it allows incorporating very realistic deployment models into connectivity studies without interference~\cite{Shao2015}. If the fading is also neglected, more network properties can be analytically evaluated, e.g., expected number of connected clusters~\cite{Kwon2016}. In this paper, we are interested in \ac{V2V} communication under the impact of interference and fading. In~\cite{Alouini2016}, a modified Mat{\`e}rn hardcore type-II process is considered for the intensity of concurrent interferers per lane, and the average multi-hop packet transmission time is calculated. In \cite{Tong2016}, the \ac{1D} Mat{\`e}rn type-II process is enhanced with discrete marks modeling non-saturated data traffic, and the transmission success probability is evaluated. In~\cite{Hourani2018}, it is shown that with low transmission probability, the outage due to \ac{1D} Bernoulli lattice converges to that due to a \ac{PPP} of equal intensity. 

While the studies~\cite{Alouini2016,Tong2016,Hourani2018} are pertinent to motorways and incorporate higher layer features (multi-hop transmission, queueing and application to automotive radars respectively), they calculate the average of the performance indicator. To the best of our knowledge, the meta distribution of the \ac{SIR} in \acp{VANET} has so far been studied only in~\cite{Abdulla2017}. Assuming a regular grid of roads and \ac{1D} \acp{PPP} for the vehicles, this (simulation-based) study indicates that the meta distribution is bimodal. Intuitively, the \ac{V2V} communication in line-of-sight experiences much higher reliability than that over an intersection, making the performance of a randomly selected link either extremely reliable or totally unreliable. 

Unlike the urban scenario in~\cite{Abdulla2017}, we would like to shed some light on the meta distribution of the \ac{SIR} along motorways. Naturally, the drivers maintain large safety distances in motorways, and the \ac{PPP} may not model accurately the locations of vehicles~\cite{Koufos2019}. In order to maintain some degree of analytical tractability while introducing more realistic deployment, we have adopted the shifted-exponential distribution for the inter-vehicle distance in~\cite{Koufos2019,Koufos2018,Koufos2019b}. This distribution has roots in transportation research~\cite{Cowan1975}, and it has also been used to model accidents for vehicles on the same lane~\cite{Sou2013}. More complex headway models like the log-normal distribution for multi-lane traffic in~\cite{Wisitpongphan2007} are difficult to analyze. The \ac{PGFL} of shifted-exponential (or hardcore) point process required to calculate the moments of the meta distribution is unknown. In order to approximate the outage probability, we calculated the moments of interference under Palm (and reduced Palm) measure \ac{w.r.t.} the shift (or hardcore distance) in~\cite{Koufos2019,Koufos2018,Koufos2019b}. Then, we selected suitable distributions for the interference level. While this approach gave good predictions for the outage probability due to the hardcore process, it is not straightforward to extend it to calculate meta distributions. 
 
Instead of pursuing interference modeling, we will deal directly with the \ac{PGFL} of the hardcore point process. Unfortunately, the bounds using first-order factorial moment expansion for Gibbs processes with conditional Papangelou intensity, see~\cite[Thereom 1]{Stucki2014}, are not tight in the upper tail of the \ac{SIR} \ac{CDF}. In order to approximate the \ac{PGFL}, we will split the contributions into near- and far-field. For the far-field, we model the interferers with a \ac{PPP}. For the near-field, we discretize the lane into intervals equal to the hardcore distance, and we allow at most one vehicle per interval. Let us call this model, \emph{the discretization model}. The main contributions of this paper are:
\begin{itemize}
\item  Using the discretization model, we devise accurate approximations for the conditional \ac{PGFL} and the meta distribution of the \ac{SIR} due to the hardcore point process. Furthermore, we illustrate that the hardcore process (and subsequently the discretization model) approximate well the meta distribution generated from synthetic motorway traces~\cite{Fiore2014,Fiore2015}, while the conventional \ac{PPP} fails. 
\item We show that introducing hardcore distance, while keeping the intensity of vehicles fixed, reduces the \ac{CoV} of the meta distribution. As a result, the conventional \ac{PPP} predicts larger disparity in the performance of different links along the motorway, and it incurs large errors in the estimation of the meta distribution generated from the traces. 
\item We show that the \ac{CoV} of the meta distribution increases for increasing \ac{SIR} threshold, while all other parameters remain fixed. As a result, the calculation of the meta distribution becomes particularly meaningful in the upper tail of the \ac{SIR} \ac{CDF}.
\end{itemize}

In Section~\ref{sec:SystemModel}, we present the system model and the discretized approximation to the hardcore process. In Section~\ref{sec:PGFL}, we calculate the \ac{PGFL} for the discretization model and in Section~\ref{sec:Meta} its meta distribution. In Section~\ref{sec:Properties}, we devise simple approximations for the first two moments of the meta distribution. In Section~\ref{sec:Validation}, we validate the models against real traces. Finally, in Section~\ref{sec:Conclusions}, we summarize the main findings and outline relevant topics for future work. 

\section{System model}
\label{sec:SystemModel}
\begin{figure*}[!t]
 \centering \subfloat[Deployment model (hardcore point process)] {\includegraphics[width=3.25in]{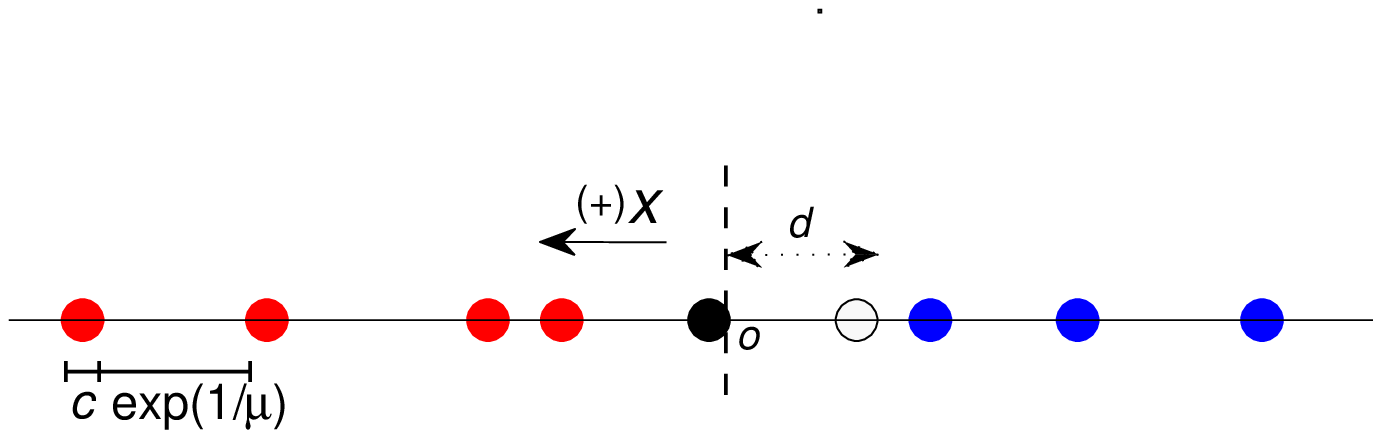}\label{fig:SystModelBeam}}\hfil \subfloat[Discretization model]{\includegraphics[width=3.75in]{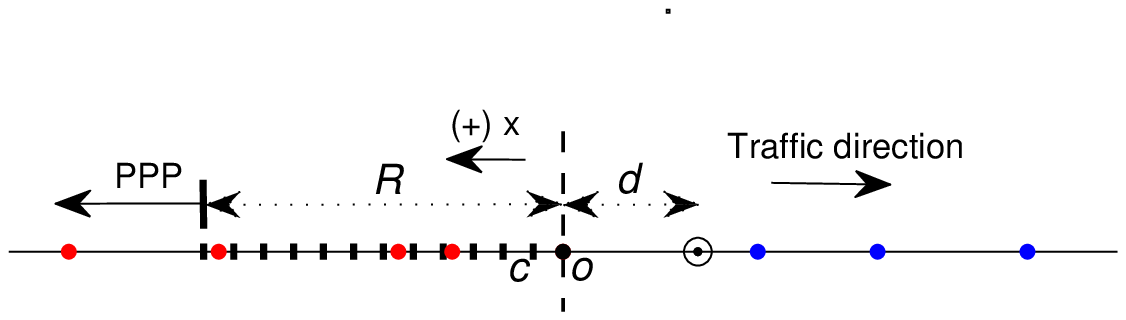}\label{fig:SystModelBeamSeg}}
 \caption{(a) The vehicles are modeled as identical impenetrable disks of diameter $c$. Their antenna is placed at the right side of the disk. A transmitter (black disk) is conditioned at the origin and paired with the receiver (hollow  disk) at $x\!=\!-d$. The vehicles behind the transmitter (red disks) generate interference at the receiver, while the rest (blue disks) do not. All vehicles move rightwards but the interferers are assumed in the positive half-axis to simplify the expressions. (b) The line segment $c\!\leq\!x\!\leq\!R$ is discretized with interval equal to $c$. The vehicles are modeled by dimensionless points located uniformly inside the discretization intervals. For $x\!>\!R$ the locations of vehicles are approximated by \ac{PPP}.}
 \label{fig:SystModel}
\end{figure*}

We consider \ac{1D} point process of vehicles $\Phi_c$, where the inter-vehicle distance follows the shifted-exponential \ac{PDF}. The shift is denoted by $c\!>\!0$ and describes the minimum safety distance from the vehicle ahead plus the average size of a vehicle. The parameter of the exponential part is denoted by $\mu\!>\! 0$ and describes the random part of inter-vehicle spacing depending on the driver's reaction time, speed, different sizes for the vehicles etc. The intensity of vehicles is $\lambda^{-1}\!=\!c\!+\!\mu^{-1}$, or $\lambda\!=\!\frac{\mu}{1+\mu c}$. We condition on the location of a transmitter at the origin. The receiver associated to it is the nearest vehicle ahead, at distance $d$, see Fig.~\ref{fig:SystModelBeam}. We assume that only the vehicles behind the transmitter generate interference. Other vehicles may also interfere due to antenna backlobes radiation, but this would not dominate the interference level, and it is currently neglected. Hereafter, the process $\Phi_c$ denotes the points with non-negative coordinate, see Fig.~\ref{fig:SystModelBeam}. 

The probability to find a vehicle at $x\!=\!r\!>\!0$ follows from the \ac{PCF}, $\rho^{\left(2\right)}\!\left(r\right)\!=\!\sum\nolimits_{k=1}^\infty \rho_k^{\left(2\right)}\!\left(r\right)$, 
\begin{equation}
\label{eq:rho2}
\rho_k^{\left(2\right)}\!\!\left(r\right) \!=\! \Bigg\{ \!\!\!\! \begin{array}{ccl} \sum\limits_{j=1}^k \!\! \frac{\mu^j \left(r-jc\right)^{j-1}}{\Gamma\left(j\right) e^{\mu\left(r-jc\right)}}, \!\!\!\!\!\!& &\!\!\!\!\!\! r\!\in\!\left(kc, \left(k\!+\!1\right)\!c\right) \\ 0, \!\!\!\!\!\!& &\!\!\!\!\!\! \text{otherwise}, \end{array} 
\end{equation}
$k\!\geq\! 1$ and $\Gamma\!\left(j\right)\!=\!\left(j\!-\!1\right)!$~\cite[equation~(32)]{Salsburg1953}.

The transmit power level is normalized to unity. The distance-based pathloss follows power-law, $r^{-\eta}$, with exponent $\eta\!>\! 2$. The fading power level is \ac{i.i.d.} over different links, following the  exponential \ac{PDF} with mean unity. Each interferer is active with probability $\xi$, independently of the activity of others.

We will now describe our novel \emph{discretization model} which will be used to approximate the \ac{CDF} and the meta distribution of the \ac{SIR} due to the hardcore process. The model splits the interferers into near- and far-field depending on their locations, see Fig.~\ref{fig:SystModelBeamSeg}. The separation threshold is denoted by $R$. The locations of vehicles for $x\!>\! R$ are approximated by a \ac{PPP} $\Phi_p$ of intensity $\lambda$, because these vehicles do not dominate the interference statistics. On the other hand, the approximating distribution for the near-field interferers considers some of the deployment contraints introduced by $\Phi_c$: Firstly, we discretize the line segment $x\!\in\!\left[c,R\right]$ into intervals of length $c$, where $R\!=\! Kc, K\!\in\!\mathbb{N}_+$. Secondly, taking into account that the minimum distance separation between successive vehicles is $c$, we allow at most one vehicle inside each interval. We assume that whether an interval contains a vehicle or not is independent of other intervals. Even though this approximation may not satisfy the hardcore constraint for all vehicles, it will suffice to approximate well the \ac{PGFL} of the point process $\Phi_c$ for realistic parameter settings. Furthermore, while the \ac{PDF} of the location of a vehicle inside the $k$-th interval, $\left(kc,\left(k\!+\!1\right)c\right)$, is available from the \ac{PCF}, we approximate it by the uniform distribution $U_k \, k\!\in\!\left\{1,2,\ldots,K\!-\!1\right\}$, to reduce the computational complexity at the cost of small accuracy loss.

Let us denote by $P_k$ the Bernoulli-distributed \ac{RV} with parameter $p_k$, equal to the probability that the $k$-th interval contains a vehicle. The parameter $p_k$ can be calculated as the integral of the \ac{PCF} within $\left[kc,\left(k\!+\!1\right)c\right]$. For instance, for $k\!=\!1$, we have $p_1\!=\!\int_c^{2c}\mu e^{-\mu\left(r-c\right)}{\rm d}r\!=\!1\!-\!e^{-c\mu}$. For large $k$, the following simplification might be of use
\begin{eqnarray}
p_k \!\!\!\!\! &=& \!\!\!\!\! \displaystyle \int_{k c}^{\left(k+1\right)c} \!\!\!\! \rho_k^{\left(2\right)}\!\!\left(r\right) {\rm d}r \nonumber \\ \!\!\!\!\! &\stackrel{(a)}{=}& \!\!\!\!\! \displaystyle \sum_{j=1}^k \frac{\Gamma\left(j,c\mu\left(k\!-\!j\right)\right)-\Gamma\left(j,c\mu\left(k\!+\!1\!-\!j\right)\right)}{\Gamma\left(j\right)} \label{eq:Pk} \\  \!\!\!\!\! &\stackrel{(b)}{\approx}& \!\!\!\!\! \displaystyle \left(k\!-\!\frac{c\mu k}{1\!+\!c\mu}\right)-\left(k\!-\!\frac{c\mu\left(k\!+\!1\right)}{1\!+\!c\mu}\right) = \lambda c, \label{eq:Pk2}
\end{eqnarray}
where $\Gamma\!\left(a,x\right)\!=\!\int_x^\infty \!\frac{t^{a-1}}{e^t}{\rm d}t$ is the incomplete Gamma function, $(a)$ follows after substituting equation~\eqref{eq:rho2} and inter-changing the order of integration and summation, and $(b)$ uses that for large $k$ the function $t\!\left(x\right)\!=\!\frac{\Gamma\left(x,c\mu\left(k-x\right)\right)}{\Gamma\left(x\right)}$ is negligible for small $x$ and ramps up to unity after some point $x_0$. Therefore we may approximate the sum $\sum_{j=1}^k\! t\!\left(j\right)$ by the integral of a unit pulse between $x_0$ and $k$. Without introducing much error in the calculation of the integral, we define $x_0$ to be the point where $t\!\left(x_0\right)\!=\!\frac{1}{2}$. In order to approximate $x_0$, we know that $\lim\limits_{x\rightarrow\infty}\frac{\Gamma\!\left(x,x\right)}{\Gamma\!\left(x\right)}\!=\!\frac{1}{2}$~\cite{Gautschi1998}. We use this property and solve for $x_0\!=\!c\mu\left(k\!-\!x_0\right)$ or $x_0\!=\!\frac{c\mu k}{1+c\mu}$ in the first term and $x_0\!=\!\!\frac{c\mu\left(k+1\right)}{1+c\mu}$ in the second, and the result follows.

\section{Approximating the conditional PGFL of $\Phi_c$}
\label{sec:PGFL}
Under \ac{i.i.d.} Rayleigh fading, the outage probability,  ${\text{P}}_{\text{out}}\!\left(\theta\right)\!=\!\mathbb{P}\left(\text{SIR}\!\leq\!\theta\right)$ is, see also~\cite[Theorem~1]{Andrews2011},  
\[
{\text{P}}_{\text{out}}\!\left(\theta\right) \!=\! 1\!-\!\mathbb{E}\Big\{\prod\limits_{x_{\!k}\in\Phi_c \!\backslash\! \left\{o\right\}} \!\!  \frac{1}{1\!+\!\xi_k s\!\left(\theta\right) \left(x_k\!+\!d\right)^{-\eta}}\Big\}, 
\]
where $s\!\equiv\!s\!\left(\theta\right)\!=\!\theta d^\eta$, and the \acp{RV} $\xi_k$ describe the activity of the $k$-th vehicle with probability $\xi$.

\noindent 
The \acp{RV} $\xi_k$ are \ac{i.i.d.} Bernoulli, and thus 
\begin{equation}
\label{eq:Pout}
{\text{P}}_{\text{out}}\!\left(\theta\right) \!=\! 1\!-\!\mathbb{E}\Big\{\!\!\prod\limits_{x_{\!k}\in\Phi_c \!\backslash\! \left\{o\right\}} \!\!\!\!\!\! \Big(1\!-\!\xi+\xi\!\! \left(1\!+\!s \left(x_k\!+\!d\right)^{-\eta}\right)^{-1}\Big)\!\!\Big\}. 
\end{equation}

The expectation in~\eqref{eq:Pout} should be taken over the locations of interferers and transmitter. The product is a \ac{RV} describing the probability of successful reception conditioned on the locations. Its distribution is essentially the meta distribution of the \ac{SIR}, as we will discuss in the next section.

After splitting the contributions to the \ac{PGFL} into near- and far-field terms, and using the discretization model, we have 
\begin{equation}
\label{eq:Outage}
\begin{array}{ccl}
{\text{P}}_{\text{out}}\!\left(\theta\right) \!\approx\! 1\!-\!\mathbb{E}\!\left\{\prod\nolimits_{x_{\!k}\in U_k} \! G_{\!n}\!\left(x_k\right) \right\} \! \mathbb{E}\!\left\{\prod\nolimits_{x_{\!k} \in \Phi_p} \!\! G_{\!f}\!\left(x_k\right) \right\}\!,
\end{array}
\end{equation}
where $G_{\!n}\!\left(x_k\right)\!=\!1-\xi p_k\!+\!\xi p_k(1\!+\!s (x_k\!+\!d)^{-\eta})^{-1}$ for the near-field and  $G_{\!f}\!\left(x_k\right)\!=\!1\!-\!\xi\!+\!\xi(1\!+\!s (x_k\!+\!d)^{-\eta})^{-1}$ for the far-field. For the near-field we have scaled the activity $\xi$ with the probability the $k$-th interval contains a vehicle. This is valid because the \acp{RV} $P_k$ are independent of each other and of $\xi_k$. 

The expectation over the far-field is straightforward to compute from the \ac{PGFL} of \ac{PPP} within $\left(R,\infty\right)$. 
\begin{equation}
\label{eq:FarField}
\begin{array}{ccl}
\mathbb{E}\left\{\prod\nolimits_{x_k\in \Phi_p} \!\! G_f\!\left(x_k\right) \right\} \!\!\!\!\! &=& \!\!\!\!\! \displaystyle  \exp\left(-\lambda\xi\!\! \int_{R+d}^\infty \frac{s}{s\!+\!x^\eta}{\rm d}x \right),
\end{array}
\end{equation}
where the integral can be expressed in terms of the hypergeometric ${}_2F_1$ function~\cite[p.~556]{Abramo}. 

The expectation for the near-field, $J_n\!=\!\mathbb{E}\!\left\{\prod\nolimits_{x_{\!k}\in U_k} \!\!G_{\!n}\!\left(x_k\right) \right\}$ requires to average over a uniform distribution for the location of a vehicle inside each interval. After bringing the expectation operator inside the product we have 
\begin{eqnarray}
J_n \!\!\!\!\! &=& \!\!\!\!\! \displaystyle  \prod\limits_{k=1}^{K\!-\!1} \frac{1}{c} \int_0^c \left(1\!-\!\xi p_k\!+\!\frac{\xi p_k}{1\!+\!s \left(x\!+\!a_k\right)^{-\eta}}\right) {\rm d}x \label{eq:NearField} \\ \!\!\!\!\! &=& \!\!\!\!\! \displaystyle  \prod\limits_{k=1}^{K\!-\!1} \!\! \left(\!1\!-\!\xi p_k\!+\!\frac{\xi p_k}{cs\left(\eta\!+\!1\right)} \left(f\!\left(a_{k+1}\right)-f\!\left(a_k\right)\right) \right), \label{eq:NearField2}
\end{eqnarray}
where $a_k\!=\!d\!+\!ck$,  $f\!\left(a_k\right)\!=\!a_k^{\eta+1}{}_2F_1\!\left(1,\!\frac{\eta+1}{\eta},\!\frac{\eta+2}{\eta};\!\frac{a_k^\eta}{-s}\right)$ and $p_k$ is given in~\eqref{eq:Pk}.

After substituting~\eqref{eq:FarField} and~\eqref{eq:NearField2} into~\eqref{eq:Outage}, and average over the link distance $d$, we get an approximation for the outage probability due to the hardcore process $\Phi_c$. We will compare the discretization model with a few other models. The first one uses a \ac{PPP} with intensity $\lambda$ in $\left(c,\infty\right)$ for the interferers and a shifted-exponential \ac{PDF} for the link distance. Hereafter, we will refer to it as model M1. The outage probability is
\begin{eqnarray}
\label{eq:ApproximationsOutage} 
{\text{P}}_{\text{out}}^{\text{M1}}\!\left(\theta\right) \!\!\!\!\! &=& \!\!\!\!\! \displaystyle  1\!-\!\int\nolimits_c^\infty \!\!\!\!  \exp\!\!\left(-\lambda\xi \!\! \int_{c+r}^\infty \frac{s\, {\rm d}x}{s\!+\!x^\eta} \right) \mu  e^{-\mu\left(r-c\right)} \! {\rm d}r.
\end{eqnarray}

The second model, model M2, assumes exponential \ac{PDF} for the inter-vehicle distances. This is the model used in the literature for linear \acp{VANET}. The outage probability is 
\begin{equation}
\label{eq:PPP2}
\begin{array}{ccl}
{\text{P}}_{\text{out}}^{\text{M2}}\!\!\left(\theta\right) \!\!\!\!\! &=& \!\!\!\!\! \displaystyle 1\!-\!\int\nolimits_0^\infty \!\!\!\! e^{\!-\lambda\xi\!\int_r^\infty \!\! \frac{\theta r^\eta x^{-\eta} {\rm d}x}{1+\theta r^\eta x^{-\eta}}} \lambda  e^{-\lambda r} {\rm d}r \\ \!\!\!\!\! &=& \!\!\!\!\!   \frac{b\left(\theta\right)}{b\left(\theta\right) + 1},  \,\, b\!\left(\theta\right)\!=\! \frac{{}_2F_1\!\left(1,1\!-\!\frac{1}{\eta},2\!-\!\frac{1}{\eta},-\theta\right)}{\left(\xi \theta\right)^{-1}\left(\eta-1\right)}.
\end{array}
\end{equation}

Equation~\eqref{eq:PPP2} is independent of the intensity. This agrees with the outage probability in the downlink of \ac{PPP} interference-limited cellular networks with nearest base station association~\cite[Eq.~(14)]{Andrews2011}. The third model M3 uses a non-homogeneous \ac{PPP} with intensity $\rho^{\left(2\right)}\!\left(x\right), x\!\geq\! 0$ and a shifted-exponential \ac{PDF} for the link distance. The outage probability follows from the \ac{PGFL} of \ac{PPP}.
\begin{eqnarray}
{} \!\!\! & & \!\!\!\!\!\!\!\!\!\!\!\!\!\!\!\!\!\!\!\!\!\!\!\! {\text{P}}_{\text{out}}^{\text{M3}}\!\left(\theta\right) \nonumber \\ {} \!\!\!\!\!\!\!\!\!\! =  &1& \displaystyle \!\!\!\!-\!\int\nolimits_c^\infty \!\!\!\!\!\! \exp\!\! \left(\!-\!\int_r^\infty \! \frac{\xi s \rho^{\left(2\right)}\!\!\left(x\!-\!d\right) {\rm d}x}{s\!+\!x^\eta} \right) \! \mu  e^{-\mu\left(r-c\right)}  {\rm d}r.  \label{eq:ApproximationsOutage2}
\end{eqnarray}

In order to calculate~\eqref{eq:ApproximationsOutage2}, we use the exact value of the \ac{PCF} for distance separation up to $4c$, and the approximation $\rho^{\left(2\right)}\!\left(x\right)\!\approx\!\lambda$ for $\left(x\!-\!d\right)\!>\!4c$. This approximation is valid for $\lambda c\!\leq\!\frac{1}{2}$, because the correlation starts to vanish beyond that distance, see~\cite[Fig.~3]{Koufos2018}. Also, $\lambda c\!\leq\!\frac{1}{2}$ is reasonable for motorways~\cite[Fig.~8 and Fig.~9]{Koufos2019}. Finally, another approximation, model M4,  yielding a closed-form expression uses a \ac{PPP} of intensity $\lambda$ in $\left(0,\infty\right)$ for the interferers and a shifted-exponential \ac{PDF} for the link.
\begin{equation}
\label{eq:PPP4}
\begin{array}{ccl}
{\text{P}}_{\text{out}}^{\text{M4}}\!\!\left(\theta\right) \!\!\!\!\! &=& \!\!\!\!\!  1 \!-\! \frac{ e^{-\tilde{b}\left(\theta\right) \mu c}}{\tilde{b}\left(\theta\right) + 1},  \,\, \tilde{b}\!\left(\theta\right)\!=\! \frac{{}_2F_1\!\left(1,1\!-\!\frac{1}{\eta},2\!-\!\frac{1}{\eta},-\theta\right)}{\left(\xi \theta\right)^{-1}\left(1+\mu c\right)\left(\eta-1\right)}.
\end{array}
\end{equation}

\begin{figure}[!t]
 \centering
  \includegraphics[width=3in]{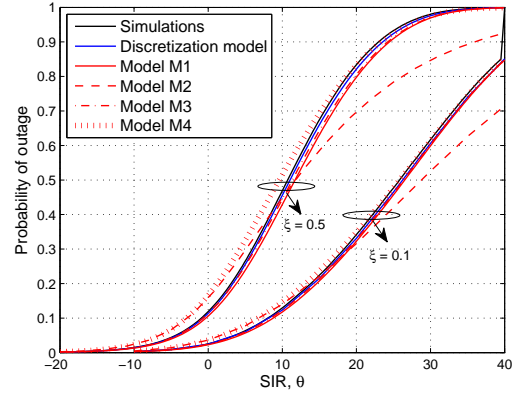}
 \caption{Simulated probability of outage due to a hardcore point process $\Phi_c$ with $\lambda\!=\! 0.025 {\text{m}}^{-1}, c\!=\!16$ m and approximations. Pathloss exponent $\eta\!=\!3$, $10^6$ simulation runs. In the discretization model, $R_{min}\!\approx\! 500$ m for $q\!=\!2\%$, see~\eqref{eq:ThresholdR}. We set $R\!=\! 512$ m, or $K\!=\!32$ units. In each simulation run: (i) We generate samples from the shifted-exponential distribution to cover a linear segment of $25$ km behind the transmitter. (ii) We generate the transmitter-receiver link distance $d$ according to the same distribution, and (iii) we generate exponentially-distributed fading and Bernoulli-distributed activity samples for all vehicles. The transmitter is assumed to be always active.}
 \label{fig:AveDOwnLaneBoth}
\end{figure}

The discretization model is a tight approximation to the simulations due to the hardcore process $\Phi_c$, see Fig.~\ref{fig:AveDOwnLaneBoth}. The approximations in~\eqref{eq:ApproximationsOutage} and~\eqref{eq:ApproximationsOutage2} perform slightly worse. The accuracy of~\eqref{eq:PPP4} is poor, unless the transmission probability is low. This corroborates the fact that the \ac{PPP} is inadequate to model interference in motorway \acp{VANET}, even if the distance distribution of the transmitter-receiver link is tuned to avoid small headways. Finally, the model M2 in~\eqref{eq:PPP2} predicts low outage for high $\theta$, because it allows unrealistically high link gains with high probability for the useful link.

In order to set the separation threshold $R$, we constrain the mean interference beyond $R$, $\mathbb{E}^{!o}\!\left\{\mathcal{I}\right\}_{x>R}$, to be less than $q\%$ of the total mean interference $\mathbb{E}^{!o}\!\left\{\mathcal{I}\right\}$. For $q\!\ll\!1$, $R$ is expected to be large, and the mean $\mathbb{E}^{!o}\!\left\{\mathcal{I}\right\}_{x>R}$ can be well-approximated using a \ac{PPP} of intensity $\lambda$. After averaging the term $\frac{\lambda\xi}{\eta-1}\left(d\!+\!R\right)^{1-\eta}$ over the link distance $d$ we get 
\[ 
\begin{array}{ccl}
\mathbb{E}^{!o}\!\left\{\mathcal{I}\right\}_{x>R} \!\!\!\!\! &\approx& \!\!\!\!\! \displaystyle \frac{\lambda \xi}{\eta-1} \int_c^\infty \left(r\!+\!R\right)^{1-\eta} \mu e^{-\mu\left(r-c\right)} {\rm d}r \\ \!\!\!\!\! &=& \!\!\!\!\! \displaystyle \frac{\lambda \xi \mu^{\eta-1}}{\eta-1} e^{\mu\left(c+R\right)} \, \Gamma\!\left(2\!-\!\eta,\mu\left(c\!+\!R\right)\right). 
\end{array}
\]

The calculation of $\mathbb{E}^{!o}\!\left\{\mathcal{I}\right\}$ involves the \ac{PCF} in~\eqref{eq:rho2} which has a complicated form. Due to the fact that $\rho^{\left(2\right)}\!\left(x\right)\!\geq\! \mu e^{-\mu c}$ for $x\!\geq\! c$~\footnote{The minimum value of $\rho^{\left(2\right)}\!\left(x\right)$ for $x\!\geq\!c$ occurs at $x\!=\!2c$, see equation~\eqref{eq:rho2} and~\cite[Fig.~3]{Koufos2018} for an illustration. This bound on the \ac{PCF} and thus on the mean interference is not tight for very regular point processes, e.g.,  $\lambda c\!>\!\frac{1}{2}$.}, the mean interference can be bounded as  
\begin{equation}
\label{eq:BoundEIo}
\begin{array}{ccl}
\mathbb{E}^{!o}\!\left\{\mathcal{I}\right\} \!\!\!\!\! &=& \!\!\!\!\! \displaystyle \xi\!\! \int_c^\infty\!\!\!\! \int_c^\infty\!\!\!\! \left(r\!+\!x\right)^{-\eta} \!\! \rho^{\left(2\right)}\!\left(x\right) \mu e^{-\mu\left(r-c\right)}{\rm d}x {\rm d}r \\  \!\!\!\!\! &\geq& \!\!\!\!\! \displaystyle \xi \!\! \int_c^\infty \!\! \frac{\mu e^{-\mu c}\left(r\!+\!c\right)^{1-\eta}}{\eta-1} \mu e^{-\mu\left(r-c\right)} {\rm d}r \\ \!\!\!\!\! &=& \!\!\!\!\! \displaystyle \frac{\xi \mu^\eta}{\eta-1} e^{\mu c} \, \Gamma\!\left(2\!-\!\eta,2\mu c\right).
\end{array}
\end{equation}

We would like to find the minimum $R$, denoted by $R^*$, satisfying $\mathbb{E}^{!o}\!\left\{\mathcal{I}\right\}_{x>R}\leq q \mathbb{E}^{!o}\!\left\{\mathcal{I}\right\}$. Since $\mathbb{E}^{!o}\!\left\{\mathcal{I}\right\}_{x>R}$ decreases in $R$, we may use the bound in~\eqref{eq:BoundEIo} instead of $\mathbb{E}^{!o}\!\left\{\mathcal{I}\right\}$. After cancelling out common terms, we end up with 
\begin{equation}
\label{eq:R}
\min\limits_{R\!>\!R^*}\!\!\left\{\!\! e^{\mu\left(c+R\right)} \Gamma\!\left(2\!-\!\eta,\mu\left(c\!+\!R\right)\right) \!\leq\! \frac{q}{1\!-\!\lambda c} \Gamma\!\left(2\!-\!\eta,2\mu c\right)\!\!\right\}\!. 
\end{equation}

The above inequality cannot be solved in closed-form \ac{w.r.t.} $R$. Due to the fact that $R$ is expected to be much larger than $\mu^{-1}$, we expand the Gamma function of the left-hand side for large $\mu R\!\gg\! 1$ obtaining: $e^{\mu\left(c+R\right)} \, \Gamma\!\left(2\!-\!\eta,\mu\left(c\!+\!R\right)\right) \leq \left(\mu R\right)^{1-\eta}$. After substituting this in the left-hand side of~\eqref{eq:R}, we get a bound $R_{min}$ on the threshold, $R\!\geq\!R_{min}\!\geq\!R^*$, satisfying the constraint $\mathbb{E}^{!o}\!\left\{\mathcal{I}\right\}_{x>R}\leq q \mathbb{E}^{!o}\!\left\{\mathcal{I}\right\}$, where 
\begin{equation}
\label{eq:ThresholdR}
R_{min} = \frac{1}{\mu} \left( \frac{q}{1\!-\!\lambda c}\, \Gamma\!\left(2\!-\!\eta,2\mu c\right)\right)^{\frac{1}{1-\eta}}.
\end{equation}

Let us assume that we increase the intensity of the hardcore process and at the same time we impose stronger thinning, keeping constant the intensity of retained transmitters $\lambda\xi$. Under this transformation, the hardcore process converges in distribution to a \ac{PPP}~\cite{Haenggi2016}. Simply thinning the process maintains some degree of correlation. Nevertheless, it is natural to assume that the strongly thinned process generates interference statistics similar to those of the thinned \ac{PPP}. In other words, a \ac{PPP} with intensity $\lambda$ in $\left(c,\infty\right)$ predicts more accurately the interference field due to the process $\Phi_c$ of equal intensity for smaller $\xi$~\cite{Hourani2018}, e.g., compare the accuracy of model M1 for $\xi\!=\!0.1$ and $\xi\!=\!0.5$ in Fig.~\ref{fig:AveDOwnLaneBoth}. Next, we show that the discretization model is consistent with this property. %for low activity, its \ac{PGFL} converges to that due to a \ac{PPP}. 
\begin{lemma}
\label{lemma:1}
For low activity $\xi$ such that $\lambda\xi\left(R\!-\!c\right)\!\ll\! 1$ and $\lambda c\!\ll\! 1$, the outage probability due to the discretization model converges to that of a \ac{PPP} given in~\eqref{eq:ApproximationsOutage} for all thresholds $\theta$.
\begin{proof}
For small $\xi$, equation~\eqref{eq:NearField} can be approximated as 
\[
\begin{array}{ccl}
J_n \!\!\!\!\! &=& \!\!\!\!\! \displaystyle  \prod\limits_{k=1}^{K\!-\!1}  \!\!\left(\!1 \!-\! \frac{p_k}{c}\!\! \int_0^c \!\!\!\! \frac{\xi s\, {\rm d}x}{s\!+\!\left(x\!+\!a_k\right)^\eta}\!\!\right) \!\stackrel{(a)}{\approx}\! \prod\limits_{k=1}^{K\!-\!1} \!\! \left(\!\!1 \!- \!\!\! \int_0^c \!\! \frac{\lambda\xi s\, {\rm d}x}{s\!+\!\left(x\!+\!a_k\right)^\eta}\!\!\right) \\ \!\!\!\!\! &\stackrel{(b)}{\gtrapprox}& \!\!\!\!\! \displaystyle 1\!-\!\sum_{k=1}^{K\!-\!1}\!\!\int_0^c\!\! \frac{\lambda \xi s \, {\rm d}x}{s\!+\!\left(x\!+\!a_k\right)^\eta} \stackrel{(c)}{\approx} \exp\!\left(\!\!-\sum_{k=1}^{K\!-\!1}\!\!\int_0^c\!\! \frac{\lambda \xi s \, {\rm d}x}{s\!+\!\left(x\!+\!a_k\right)^\eta}\!\right) \\ \!\!\!\!\! &\stackrel{(d)}{=}& \!\!\!\!\!  \displaystyle \exp\!\left(\!\!-\lambda\xi\!\! \int\nolimits_{\!c+d\!}^{\!R+d\!}\!\! \frac{s \, {\rm d}x}{s\!+\!x^\eta}\!\right)\!\!. 
\end{array}
\]

The approximation in $(a)$ follows from $p_k\!\approx\!\lambda c \, \forall k$. This is valid for large $k$, see~\eqref{eq:Pk2}, and it can also be valid for small $k$ under the condition $\lambda c\!\ll\! 1$. For instance, $p_1\!=\!1-e^{-\frac{\lambda c}{1-\lambda c}}\!=\! \lambda c\!+\!o\!\left(\lambda^2c^2\right)$. The inequality in $(b)$ is the Weierstrass product inequality $\prod_i\!\left(1\!-\!y_i\right)\!\geq\!1\!-\!\sum_i\!y_i,\, y_i\!\in\!\left[0,1\right]$, and $(c)$ follows from the expansion $e^{-x}\!\approx\!1-x, x\!\rightarrow\! 0$. These approximations hold for $\sum_{k=1}^{K\!-\!1}\!\!\int_0^c\!\! \frac{\lambda\xi s \, {\rm d}x}{s+\left(x+a_k\right)^\eta}\!\leq\!\lambda \xi \left(K\!-\!1\right)c \!=\!\lambda \xi \left(R\!-\!c\right)\!\ll\! 1$. We get $(d)$, which is the expression for the \ac{PGFL} of a \ac{PPP} in $\left(c,R\right)$, after adding up the $\left(K\!-\!1\right)$ integral terms. After multiplying the above simplification for $J_n$ with the far-field term in~\eqref{eq:FarField}, we end up with the desired result. 
\end{proof}
\end{lemma}

\section{The Meta Distribution of the SIR for $\Phi_c$}
\label{sec:Meta}
The meta distribution of the \ac{SIR} is the \ac{PDF} of the probability of successful reception, ${\text{P}}_s\!\left(\theta\right)$, conditioned on the spatial realization, i.e., fixed but unknown locations for the vehicles. ${\text{P}}_s\!\left(\theta\right)\!=\! \mathbb{P}\!\left({\text{SIR}}\!>\!\theta|\Phi_c\right)$. The conditional probability is computed over fast fading and ALOHA, see~\eqref{eq:Pout}. 
\begin{equation}
\label{eq:Pstheta}
{\text{P}}_s\!\left(\theta\right) = \prod\nolimits_{x_{\!k}\in\Phi_c \!\backslash\! \left\{o\right\}} \!\! \left(1\!-\!\xi\!+\!\xi\left(1\!+\!s \left(x_k\!+\!d\right)^{-\eta}\right)^{\!-1}\right). 
\end{equation}

Due to the ergodicity of the point process $\Phi_c$, the \ac{PDF} of the \ac{RV} ${\text{P}}_s\!\left(\theta\right)$ is equivalent to the spatial distribution of the probability of successful reception given a realization of the point process. The complementary \ac{CDF} $\mathbb{P}\!\left({\text{P}}_s\!\left(\theta\right)\!>\!u\right), u\!\in\!\left[0,1\right]$, indicates in each realization of $\Phi_c$, the fraction of scheduled links that experience an \ac{SIR} higher than $\theta$ with probability at least $u$. The calculation of the \ac{PDF} is not easy, but for the moments we may follow~\cite[Appendix]{Haenggi2016}. The $b$-th moment, $M_b\!\left(\theta\right)\!\equiv\!M_b$, is computed by raising~\eqref{eq:Pstheta} to the $b$-th power and taking its expectation over $x_k$ and $d$ 
\[
\begin{array}{ccl}
M_b \!\!\!\!\! &=& \!\!\!\!\! \displaystyle \mathbb{E}\Big\{\prod\nolimits_{x_{\!k}\in\Phi_c \!\backslash\! \left\{o\right\}} \!\! \left(1\!-\!\xi\!+\!\xi\left(1\!+\!s \left(x_k\!+\!d\right)^{-\eta}\right)^{\!-1}\right)^{\!b}\Big\}.
\end{array}
\]

Using the discretization model, the contribution to the moment $M_b$ from the near-field can be written as 
\[
\begin{array}{ccl}
M_{\!b,n} \!\!\!\!\!\! &=& \!\!\!\!\! \displaystyle \mathbb{E}\!\Big\{ \prod\nolimits_{x_{\!k}\in U_k} \!\! \left(\frac{1}{1\!+\!s \xi_k P_k \left(x_k\!+\!d\right)^{-\eta}}\right)^{\!b}\!\! \Big\} \\ {} \!\!\!\!\!\!\!\! &=& \!\!\!\!\! \displaystyle \mathbb{E}\!\Big\{ \prod\nolimits_{x_{\!k}\in U_k} \!\! \left(1\!-\!\xi\!+\!\frac{\xi}{1\!+\!s P_k \left(x_k\!+\!d\right)^{-\eta}}\right)^{\!b}\!\! \Big\}. 
\end{array}
\]

\noindent 
Taking the average \ac{w.r.t.} the Bernoulli \acp{RV} $P_k$ yields 
\begin{equation}
\label{eq:Mbn}
\begin{array}{ccl}
M_{\!b,n} \!\!\!\!\!\!\! &=& \!\!\!\!\! \displaystyle \mathbb{E}\!\Big\{ \!\!\!\!\! \prod\limits_{x_{\!k}\in U_k} \!\!\!\! \left(\! 1\!-\!p_k \!\!+\!\!p_k\!\! \left(\!1\!-\!\xi\!+\!\frac{\xi}{1\!\!+\!s\! \left(x_k\!+\!d\right)^{-\eta}}\!\right)^{\!\!b}\right)\!\!\Big\}. 
\end{array}
\end{equation}

Note that $M_{1,n}\!=\!J_n$, see~\eqref{eq:NearField}. Next, we generalize Lemma~\ref{lemma:1} for $b\!>\!1$. 
\begin{lemma}
\label{lemma:2}
For low activity $\xi$ such that $\lambda\xi b\left(R\!-\!c\right)\!\ll\! 1$ and $\lambda c\!\ll\! 1$, $M_{b,n} \!\approx\! \exp\!\big(\!\!-\!\!\lambda\xi\!\int_{\!c+d}^{\!R+d} 1\!-\!\big(1\!-\!\frac{\xi s}{s+x^\eta}\big)^b\!{\rm d}x\big) \,\forall \theta$.
\begin{proof}
Firstly, we take the expectation over $x_k$ for each term of the product in~\eqref{eq:Mbn} obtaining 
\begin{equation}
\label{eq:Mbn2}
M_{b,n}  \!=\! \prod\limits_{k=1}^{K\!-\!1}\!\! \left(1\!-\!p_k\!\! \left( 1 \!-\!  \frac{1}{c} \! \int_0^c \!\!\! \left( 1\!-\! \frac{\xi s}{s\!+\!\left(x\!+\!a_k\right)^\eta} \right)^{\!b} \!\! {\rm d}x \!\right) \! \right)\!. 
\end{equation}

Secondly, we follow exactly the same steps as in the proof of Lemma~\ref{lemma:1}. After substituting $p_k\!\approx\! \lambda c\, \forall k$ in $M_{b,n}$, the sufficient condition for the lemma to hold is 
\[
\begin{array}{lll} 
 \displaystyle \lambda c \sum\nolimits_{k=1}^{K\!-\!1} \! \left(1\!-\!\frac{1}{c} \!\! \int_0^c \!\! \left( 1\!-\! \frac{\xi s}{s\!+\!\left(x\!+\!a_k\right)^\eta} \right)^b \! {\rm d}x\right) &\stackrel{(a)}{\leq}& \\  \displaystyle \lambda c \sum\nolimits_{k=1}^{K\!-\!1} \left(1 \!-\! \left(1-\xi\right)^b\right) \stackrel{(b)}{\approx} \lambda c \left(K\!-\!1\right) \xi b &=& \\  \displaystyle \lambda\xi b\left(R\!-\!c\right)\!\ll\! 1, 
\end{array}
\]
where $(a)$ holds $\forall s$, and $(b)$ is true for $\xi b\!\ll\! 1$ which is met under the condition $\lambda\xi b\left(R\!-\!c\right)\!\ll\! 1$ for $\lambda\left(R\!-\!c\right)\!\geq\!1$, i.e., more than a single vehicle (on average) in the near-field.
\end{proof}
\end{lemma}

The numerical calculation of~\eqref{eq:Mbn2} can be simplified using binomial expansion in the integrand yielding 
\begin{equation}
\label{eq:MetaNear}
\begin{array}{lcl}
M_{b,n} \!\!\!\!\!\!\! &=& \!\!\!\!\!\!\! \displaystyle \prod\limits_{k=1}^{K\!-\!1} \!\!\bigg(\!\!1\!-\!p_k\!+\!p_k\!\! \sum_{j=0}^b \!\! \binom{b}{j} \frac{h\!\left(a_{k\!+\!1},\!j\right)\!-\!h\!\left(a_k,\!j\right)}{\left(-\xi\right)^{\!-j}}\!\!\bigg)\!,
\end{array}
\end{equation}
where $h\!\left(a_k,j\right)\!=\!\frac{a_k}{c}{}_2F_1\!\left(j,\!\frac{1}{\eta},\!\frac{\eta+1}{\eta};\!\frac{a_k^\eta}{-s}\right)$. 

The contribution $M_{b,f}\!=\!\mathbb{E}\!\left\{\!\prod\nolimits_{x_{\!k}\in\Phi_p} \!\! G_{\!f}\!\left(x_k\right)^b \!\right\}$, uses the \ac{PGFL} of \ac{PPP}, see~\cite[App. A]{Haenggi2013a}.
\begin{equation}
\label{eq:MetaFar}
\begin{array}{ccl}
M_{b,f} \!\!\!\!\! &=& \!\!\!\!\! \displaystyle \exp\!\left(\!-\lambda\int_{\!R+d}^{\!\infty} 1\!-\!\left(1\!-\!\frac{\xi s}{s\!+\!x^\eta}\right)^b{\rm d}x\right) \\ \!\!\!\!\! &=& \!\!\!\!\! \displaystyle \exp\!\bigg(\lambda \sum_{j=1}^b\binom{b}{j}\left(-\xi s\right)^j\underbrace{\int_{\!R+d}^{\!\infty}\!\! \frac{{\rm d}x}{\left(s\!+\!x^\eta\right)^j}}_{F_j}\bigg), 
\end{array}
\end{equation}
where $F_j\!=\!\frac{\left(R+d\right)^{1-j\eta}}{\left(j\eta-1\right)} {}_2F_1\!\left(j\!-\!\frac{1}{\eta},j,j\!-\!\frac{1}{\eta}\!+\!1;\!-s\left(R\!+\!d\right)^{-\eta} \right)$.

The moment $M_b$ is calculated by multiplying~\eqref{eq:MetaNear} with~\eqref{eq:MetaFar} before averaging over the shifted-exponential distribution for the link distance. The integration becomes computationally demanding for high $b$ hence, we will calculate only the first two moments, and match them to a Beta \ac{PDF} $g\!\left(z\right)\!=\!\frac{1}{{\text{B}}\left(\alpha,\beta\right)} z^{\alpha-1}\left(1\!-\!z\right)^{\beta-1}, \, \alpha,\beta\!>\!0$, where ${\text{B}}\!\left(\alpha,\beta\right)\!=\!\frac{\Gamma\left(\alpha\right) \Gamma\left(\beta\right)}{\Gamma\left(\alpha+\beta\right)}$. The Beta approximation has been widely adopted for meta distributions~\cite{Haenggi2016,Haenggi2018}. Its parameters are $\alpha = \frac{m_1\left(m_1\left(1-m_1\right)-m_2\right)}{m_2}$ and $\beta = m_1\!-\!1+\frac{m_1\left(1-m_1\right)^2}{m_2}$, where $m_1\!=\!M_{1,n}M_{1,f}$ and $m_2\!=\!M_{2,n}M_{2,f}\!-\!m_1^2$.~\cite[Sec. II-F]{Haenggi2016}.

We will also approximate the meta distribution due to the models M1 and M2. In Fig.~\ref{fig:AveDOwnLaneBoth}, we saw that the model M3 involves much more complicated integration than M1 without obvious benefits. The model M4 improves the prediction as compared to M2 but not as much as M1. For the model M1 the moments $M_b$ are calculated after substituting $R\!=\!c$ in~\eqref{eq:MetaFar} and averaging over $\mu e^{-\mu\left(r-c\right)}, r\!>\!c$. For the model M2 the moments are computed after substituting $R\!=\!0$ in~\eqref{eq:MetaFar} and averaging over $\lambda e^{-\lambda r}, r\!>\!0$. For all models we will apply the method of moments with Beta approximation.
\begin{figure}[!t]
 \centering
  \includegraphics[width=3in]{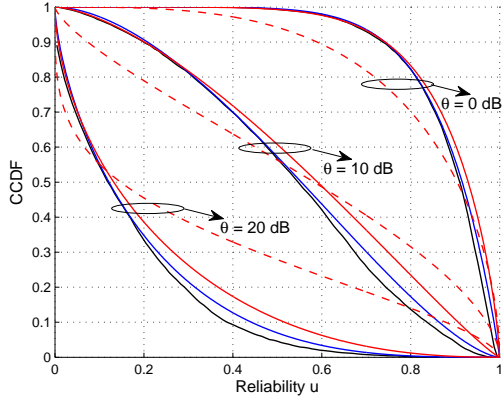}
 \caption{The simulated meta distribution of the \ac{SIR} for a hardcore process with intensity $\lambda\!=\!0.025 {\text{m}}^{-1}$,  $c\!=\!16 {\text{m}}$ and $\xi\!=\!0.5$ (black lines), and the Beta approximations using the discretization model (blue lines) and the models M1 (solid red lines) and M2 (dashed red lines). The simulations are generated using $10^4$ spatial configurations with $10^4$ realizations of fading and activity per configuration. See the caption of Fig.~\ref{fig:AveDOwnLaneBoth} for other parameter settings.}
 \label{fig:MetaDiscrPPPBetaApprox}
\end{figure}

\begin{figure}[!t]
 \centering
  \includegraphics[width=3in]{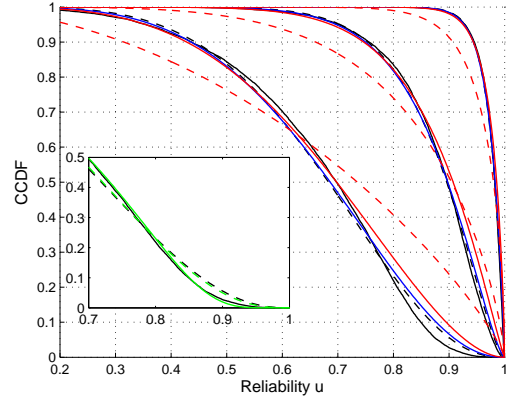}
 \caption{The meta distribution of the \ac{SIR} for a hardcore process with activity $\xi\!=\!0.1$. See the caption of Fig.~\ref{fig:MetaDiscrPPPBetaApprox} for other parameter settings and explanation of line styles. The black dashed lines correspond to a Beta approximation using the simulated mean and variance due to the hardcore process. This is to illustrate that for low activity and large thresholds $\theta$, the Beta approximation may not be very accurate at the tail. In the inset, we illustrate that for threshold $20$ dB, a generalized Beta \ac{PDF} with two parameters (green dashed line), $g\!\left(z\right)\!=\!\alpha\beta z^{\alpha-1}\left(1\!-\!z^\alpha\right)^{\beta-1}$, gives a marginal improvement, and with three parameters (green solid line), $g\!\left(z\right)\!=\!\frac{\alpha}{{\text{B}}\left(\epsilon,\beta\right)} z^{\alpha\epsilon-1}\left(1\!-\!z^\alpha\right)^{\beta-1}$, matching also the skewness, gives almost a perfect fit. The penalty paid in both cases is the numerical iterative calculation of their parameters.}
 \label{fig:MetaDiscrPPPBetaApprox01}
\end{figure}

The simulated meta distribution due to the hardcore process and the various approximations are depicted in Fig.~\ref{fig:MetaDiscrPPPBetaApprox}. The first point to remark is that M2 can be very unreliable, especially for large thresholds, $-$ it significantly overestimates both moments, see also Fig.~\ref{fig:AveDOwnLaneBoth}. Correcting M2 using M1 already brings a major improvement, which can be enhanced further using the discretization model. Reading from the figure, M2 predicts that $70\%$ of scheduled links achieve a \ac{SIR} of $0$ dB with probability at least $0.8$ while, the other models predict very accurately the correct fraction $83\%$ of links. Another remark from Fig.~\ref{fig:MetaDiscrPPPBetaApprox} is that for low thresholds $\theta$, the \ac{CoV} of the meta distribution decreases, which means that most of the scheduled links will experience about the same reliability. The meta distribution for activity $\xi\!=\!0.1$ is depicted in Fig.~\ref{fig:MetaDiscrPPPBetaApprox01}.
\section{Properties of the meta distribution}
\label{sec:Properties}
In this section we study the behaviour of the moments $M_1, M_2$ and the \ac{CoV} of the meta distribution \ac{w.r.t.} the activity and the \ac{SIR} threshold. In addition, we devise low-complexity approximations for these terms, which turn out to be much more accurate than the predictions using the model M2. 
\begin{lemma}
\label{lemma:3}
For $\lambda c\!\ll\! 1$, the \ac{CoV} of the meta distribution for the discretization model as $\xi\!\rightarrow\! 0$,  $\theta\!\rightarrow\!\infty$ such that $\xi\theta^{\frac{1}{\eta}}\!=\!T$, where $T\!>\!0$ is constant, converges to $\frac{\nu}{\sqrt{\mu\left(\mu+2\nu\right)}}$, where $\nu\!=\!\lambda T \frac{\eta^2\!+\!1}{\eta^2\!-\!1}$. 
\begin{proof}
For $\xi\!\rightarrow\! 0$, the condition $\lambda\xi b\left(R\!-\!c\right)\!\ll\! 1, b\!\in\!\left\{1,2\right\}$ is satisfied for realistic $\lambda,c,R$. In addition $\lambda c\!\ll\! 1$ hence, the assumptions in Lemma~\ref{lemma:2} hold. Therefore the moments $M_1, M_2$ due to the discretization model can be approximated using the model M1. The first moment due to M1, see~\eqref{eq:ApproximationsOutage}, is 
\begin{equation}
\label{eq:Lemma3M1init}
M_1=\int_c^\infty e^{-\lambda\xi\int_{c+r}^\infty\frac{\theta r^\eta {\rm d}x}{\theta r^\eta + x^\eta}}\mu e^{-\mu\left(r-c\right)} {\rm d}r. 
\end{equation}

For large $\theta$, the integral \ac{w.r.t.} $x$ in~\eqref{eq:Lemma3M1init} accepts the following approximation. 
\[
\begin{array}{ccl}
\int_{c+r}^\infty\frac{\theta r^\eta {\rm d}x}{\theta r^\eta + x^\eta} \stackrel{(a)}{\approx} \int_{c+r}^{r\theta^{\frac{1}{\eta}}}\left(1-\frac{x^\eta}{\theta r^\eta}\right){\rm d}x + \int_{r\theta^{\frac{1}{\eta}}}^\infty \theta r^\eta x^{-\eta} {\rm d}x & = & \\ \frac{\eta^2\!+\!1}{\eta^2\!-\!1} r \,  \theta^{\frac{1}{\eta}}\!-\!\left(c\!+\!r\right)\!+\!\frac{\left(c+r\right)^{1+\eta}r^{-\eta}}{\left(\eta+1\right)\theta} \approx
\frac{\eta^2\!+\!1}{\eta^2\!-\!1} r \, \theta^{\frac{1}{\eta}}\!-\!\left(c\!+\!r\right), 
\end{array}
\]
where $(a)$ follows from expanding the fraction (up to first-order) for small $x\!<\!\theta^{1/\eta} r$ in the first term, and large $x\!>\!\theta^{1/\eta} r$ in the second term. 

After substituting the above approximation in~\eqref{eq:Lemma3M1init}, and carrying out the integration \ac{w.r.t.} $r$, we end up with 
\begin{equation}
\label{eq:Lem3M1}
M_1\approx e^{\lambda c \xi} \, \frac{\mu e^{-\nu_1 c}}{\mu\!+\!\nu_1}, \, \nu_1=\lambda\xi\left(\frac{\eta^2\!+\!1}{\eta^2\!-\!1}\,  \theta^{\frac{1}{\eta}}-1\right). 
\end{equation}

\noindent
Similarly, for the second moment we get
\begin{equation}
\label{eq:Lem3M2}
\begin{array}{ccl}
M_2 \!\!\!\!\! &\approx& \!\!\!\!\! e^{2 \lambda c \xi-\lambda\xi^2 c} \, \frac{\mu e^{-\nu_2 c}}{\mu+\nu_2}, \\ 
\nu_2 \!\!\!\!\! &=& \!\!\!\!\! 2\lambda\xi\!\!\left(\frac{\eta^2\!+\!1}{\eta^2\!-\!1}\,  \theta^{\frac{1}{\eta}}\!-\!1\right)\!-\!\lambda\xi^2\!\!\left(\!\frac{4\eta^3+3\eta+1}{\left(\eta+1\right)\left(4\eta^2-1\right)}\, \theta^{\frac{1}{\eta}}\!-\!1\!\right).
\end{array}
\end{equation}

After substituting $\theta^{\frac{1}{\eta}}\!=\!T\xi^{-1}$ in $\nu_1,\nu_2$ and taking the limit of $M_1, M_2$ for $\xi\!\rightarrow\!0$, we get  
\begin{equation}
\label{eq:Lem3M1M2Limits}
\lim\limits_{\xi\rightarrow 0}M_1\!=\!\frac{\mu e^{-\nu c}}{\mu+\nu},\,\,  \lim\limits_{\xi\rightarrow 0}M_2\!=\!\frac{\mu e^{-2\nu c}}{\mu+2\nu}.
\end{equation}

Finally, writing the \ac{CoV} as $\sqrt{\frac{M_2}{M_1^2}\!-\!1}$, and substituting the limits from~\eqref{eq:Lem3M1M2Limits} yields the desired result. The limits depend on $\xi, \theta$ through the product $\xi\theta^{1/\eta}$, while the approximations~\eqref{eq:Lem3M1} and~\eqref{eq:Lem3M2} depend also individually on $\xi$.
\end{proof}
\end{lemma}

\begin{figure}[!t]
 \centering
  \includegraphics[width=3in]{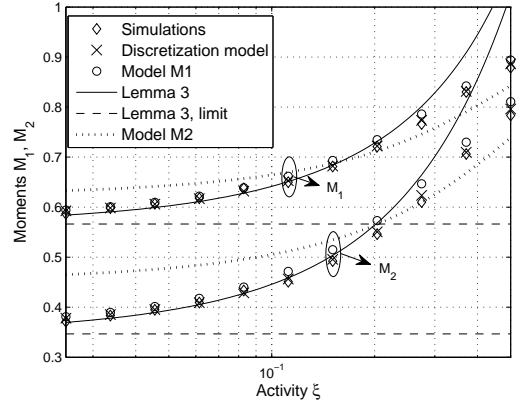}
  \caption{The moments $M_1, M_2$ for pairs $\left(\xi,\theta\right)$ satisfying $T\!=\!\xi\theta^{1/\eta}\!=\!0.5$. The discretization model, the model M1 and the approximations~\eqref{eq:Lem3M1} and~\eqref{eq:Lem3M2} in Lemma~\ref{lemma:3} estimate accurately the moments due to the simulated hardcore process. The model M2 fails. See the caption of Fig.~\ref{fig:MetaDiscrPPPBetaApprox} for other parameters.}
 \label{fig:MetaM1M2AsymptotT05}
\end{figure}

In order to illustrate the usefulness of Lemma~\ref{lemma:3}, let us assume early penetration of \acp{VANET}, with activity $\xi\!=\!10^{-2}$. Let us also consider a high-rate data communication at $\theta\!=\!30$ dB, yielding $T\!=\!0.1$ for $\eta\!=\!3$. The numerical calculation of the moments $M_1, M_2$ and the \ac{CoV} using various models are presented in Table~\ref{table1}. The model M1 is a good approximation to the discretization model because the activity is low. The approximations~\eqref{eq:Lem3M1} and~\eqref{eq:Lem3M2} follow closely the results due to M1, because the threshold $\theta$ is large. In addition, the limit ($\xi\!\rightarrow\! 0$) for the \ac{CoV} in Lemma~\ref{lemma:3} works well yielding, $\frac{\nu}{\sqrt{\mu\left(\mu+2\nu\right)}}\!=\!0.070$. Neglecting the hardcore distance, i.e., $\mu\!=\!\lambda$ and  $\frac{\nu}{\sqrt{\lambda\left(\lambda+2\nu\right)}}\!=\!0.112$ incurs large error, because the link gain experiences much higher variability for $c\!=\!0$ than for $c\!=\!16$, while $\lambda$ is fixed. In Fig.~\ref{fig:MetaM1M2AsymptotT05}, it is illustrated that the approximations in~\eqref{eq:Lem3M1} and~\eqref{eq:Lem3M2} are good also for realistic activity values, e.g., up to $\xi\!=\!0.2$. For $\xi\!=\!0.5$ we obtain $\theta\!=\!1$ for $T\!=\!0.5$, and the approximations in~\eqref{eq:Lem3M1},~\eqref{eq:Lem3M2} break down.  

\begin{figure}[!t]
 \centering
  \includegraphics[width=3in]{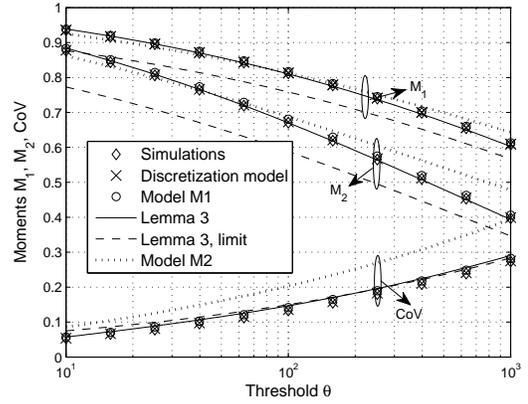}
 \caption{The moments $M_1, M_2$ and the \ac{CoV} for activity $\xi\!=\!0.05$ and large thresholds $\theta$. The discretization model, the model M1 and the approximations~\eqref{eq:Lem3M1},~\eqref{eq:Lem3M2} give very good predictions to the simulated hardcore process, while M2 fails. The limit for the \ac{CoV} in Lemma~\ref{lemma:3} increases approximately by $\theta^{1/2\eta}$ for large $\theta$. See the caption of Fig.~\ref{fig:MetaDiscrPPPBetaApprox} for other parameter settings.}
 \label{fig:MetaM1M2CoV}
\end{figure}

The limit of the \ac{CoV} in Lemma~\ref{lemma:3} increases in $\nu$ and thus in $\theta$, while $\xi$ is kept constant, see Fig.~\ref{fig:MetaM1M2CoV} for an illustration. This is also evident by visual inspection from Fig.~\ref{fig:MetaDiscrPPPBetaApprox}, where the variance of the meta distribution is much less for $\theta\!=\!1$ than for $\theta\!=\!100$. The large \acp{CoV} mean that the average success probability $M_1$ does not represent well the performance of different links across a snapshot of the network. Furthermore, Lemma~\ref{lemma:3} indicates that for low and decreasing activity $\xi\!\rightarrow\! 0$, while keeping constant $M_1$ by increasing $\theta$, the variance of the conditional success probability across the network does not become zero. This is in accordance with~\cite[Corollary~7]{Haenggi2016} for cellular networks with random activity. In Fig.~\ref{fig:MetaM1M2AsymptotT05}, for $M_1\!\approx\!0.566$ and activity $\xi\!\rightarrow\! 0$, the standard deviation of the conditional success probability converges to $0.16$. Finally, according to~\eqref{eq:Lem3M1M2Limits}, for $\eta\!>\!2$ and $\theta\!\geq\!1$,  $M_1, M_2$ increase for increasing $\eta$ but the \ac{CoV} decreases. Intuitively, higher pathloss means that links scheduled at the same time are better isolated from each other, and the fraction of links achieving certain reliability should increase, see also Fig.~\ref{fig:DataPout} and Fig.~\ref{fig:DataPout1200}.  
\begin{table}[!t]
\caption{The discretization model, the model M1, and the approximations~\eqref{eq:Lem3M1},~\eqref{eq:Lem3M2} predict well the moments $M_1,M_2$ and the \ac{CoV} of the meta distribution due to the simulated hardcore process. The model M2 fails in the estimation of \ac{CoV}. Each term is rounded at the fifth decimal digit. $\xi\!=\!10^{-2}$, $\theta\!=\! 10^3$, see the caption of Fig.~\ref{fig:MetaDiscrPPPBetaApprox} for other parameter settings.}
\label{table1}
\centering
\begin{tabular}{ |l|c|c|c|c| } 
 \hline
  {} & $M_1$ & $M_2$ & \ac{CoV} &  \\ \hline 
 Discretization model & $0.89968$ & $0.81293$ & $0.06580$ & \\  \hline 
 Model M1 & $0.90050$ & $0.81456$ & $0.06729$ & \\ \hline 
 Lemma~\ref{lemma:3},~\eqref{eq:Lem3M1} and~\eqref{eq:Lem3M2} & $0.89698$ & $0.80846$ & $0.06954$ & \\  \hline 
Model M2 & $0.90015$ & $0.81891$ & $0.10322$ & \\  \hline 
 Simulations hardcore & $0.89811$ & $0.81010$ & $0.06579$ & \\  \hline 
\end{tabular}
\end{table}

Next, we present approximations for $M_1, M_2$ and \ac{CoV} for $\theta\!\ll\!1$. For small $\theta$, the model M1 approximates well the discretization model even for large activity, see Fig.~\ref{fig:AveDOwnLaneBoth} and Fig.~\ref{fig:MetaDiscrPPPBetaApprox} for example illustrations. 
\begin{lemma}
\label{lemma:4}
For $\theta\!\ll\!1$, the \ac{CoV} of the meta distribution for the discretization model is approximated by  $\frac{\nu^*}{\sqrt{\mu\left(\mu+2\nu^*\right)}}$, where $\nu^*\!=\! \frac{\left(\eta+1\right)\lambda t}{2^\eta\left(\eta-1\right)}$ and $t\!=\!\xi\theta\!\ll\!1$ is a positive constant. 
\begin{proof}
For small $\theta$, equation~\eqref{eq:Lemma3M1init} can be approximated as 
\[
\begin{array}{ccl}
M_1 \!\!\!\!\! &\approx& \!\!\!\!\! \int_c^\infty e^{-\lambda\xi\int_{c+r}^\infty\theta r^\eta x^{-\eta} {\rm d}x}\mu e^{-\mu\left(r-c\right)} {\rm d}r \\ \!\!\!\!\! &=& \!\!\!\!\! \int_c^\infty e^{-\frac{\lambda\xi\theta}{\eta-1} r^\eta \left(r+c\right)^{1-\eta}}\mu e^{-\mu\left(r-c\right)} {\rm d}r. 
\end{array}
\]

For small $\theta$, we get $\mu\!>\!\frac{\lambda\xi\theta}{\eta-1}$, which means that the term $e^{-\mu r}$ dominates the integrand. In addition, the main contribution to the integral is given by the vicinity of $c$. Because of that, we can expand the term $r^\eta \left(r\!+\!c\right)^{1-\eta}$ around $c$, without introducing much error. After expanding up to the first order and carrying out the integration, we end up with 
\begin{equation}
\label{eq:Lem4M1}
M_1 \approx e^{\frac{\lambda c t}{2^\eta}} \, \frac{\mu e^{-\nu^* c}}{\mu+\nu^*}. 
\end{equation}

Similarly, the approximation of $M_2$ for $\theta\!\ll\!1$, keeping only the leading-order term \ac{w.r.t.} $t$, is 
\begin{equation}
\label{eq:Lem4M2}
M_2\approx \, e^{\frac{\lambda c t}{2^{\eta-1}}} \, \frac{\mu e^{-2\nu^* c}}{\mu\!+\!2\nu^*}.
\end{equation}

The approximations~\eqref{eq:Lem4M1} and~\eqref{eq:Lem4M2} depend on $\xi$ and $\theta$ only through their product.  After substituting~\eqref{eq:Lem4M1},~\eqref{eq:Lem4M2} into $\sqrt{\frac{M_2}{M_1^2}\!-\!1}$, the result of the lemma follows.
\begin{figure}[!t]
 \centering
  \includegraphics[width=3in]{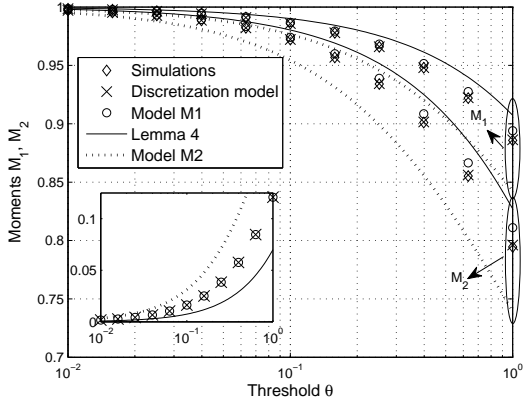}
 \caption{The moments $M_1, M_2$ for activity $\xi\!=\!0.5$ and small thresholds $\theta$. The discretization model and the model M1 predict well the simulated moments due to the hardcore process. The approximations~\eqref{eq:Lem4M1},~\eqref{eq:Lem4M2} give better predictions than M2. The \ac{CoV} (see the inset) is approximately proportional to $\theta$ for small $\theta$, see Lemma~\ref{lemma:4}. See the caption of Fig.~\ref{fig:MetaDiscrPPPBetaApprox} for other parameters.}
 \label{fig:M1M2SmallTheta}
\end{figure}
\end{proof}
\end{lemma}

The approximations~\eqref{eq:Lem4M1},~\eqref{eq:Lem4M2} are validated in Fig.~\ref{fig:M1M2SmallTheta}. Summing up, according to Lemma~\ref{lemma:3}, if we reduce the activity by $10$ dB, we can increase the threshold $\theta$ by $10\eta$ dB to maintain the same probability of success $M_1$ for large $\theta$ and $\xi\!\rightarrow\! 0$. According to Lemma~\ref{lemma:4}, for low $\theta$, we can increase the threshold only by $10$ dB. For instance, in Fig.~\ref{fig:AveDOwnLaneBoth}, the probability of outage at $-10$ dB is $0.014$ with activity $\xi\!=\!0.5$. The same outage with activity $\xi\!=\!0.1$, occurs around $-3$ dB confirming~\eqref{eq:Lem4M1}. Note also that according to~\eqref{eq:Lem4M2}, the two pairs $\left(\xi,\theta\right)\!\in\!\left\{\left(0.5,-10 \text{dB}\right),\left(0.1,-3 \text{dB}\right)\right\}$ result in the same second moment $M_2$ too. On the other hand, the probability of outage at $10$ dB is $0.45$ with activity $\xi\!=\!0.5$ in Fig.~\ref{fig:AveDOwnLaneBoth}, while with activity $\xi\!=\!0.1$ the same outage occurs at $24.5$ dB, instead of $31$ dB predicted by~\eqref{eq:Lem3M1M2Limits}. This is because~\eqref{eq:Lem3M1M2Limits} is valid for  $\xi\!\rightarrow\!0$, and~\eqref{eq:Lem3M1} should be used instead. Finally, the \acp{CoV} calculated in the Lemmas decrease in $c$, while the intensity $\lambda$ is fixed. This is clear after writing the \ac{CoV} as $\frac{\nu\mu^{\!-1}}{\sqrt{1+2\nu\mu^{\!-1}}}$, where $\nu\mu^{\!-1}\!\propto\! \left(1\!-\!\lambda c\right)$, see Fig.~\ref{fig:MetaM1M2CoV} and the inset in Fig.~\ref{fig:M1M2SmallTheta} for the comparison between M2 and the Lemmas.

\section{Validation with real traces}
\label{sec:Validation}
Thanks to~\cite{Fiore2014,Fiore2015}, there are publicly available traces for three-lane unidirectional motorway traffic. In order to generate them, the studies have collected per-lane measurements every second about the number of passing vehicles and their speed, using induction loops at a measurement location outside Madrid in Spain. The measurements have been used to calibrate a simulator modeling micro-mobility features like lane-changing pattern, velocity distributions etc. The output of the simulator is the location of vehicles, i.e., lane and horizontal position over a road segment of 10 km with one second granularity. In~\cite{Koufos2019} we have analyzed these traces using the J- and the Ripley's K-function~\cite[Ch. 2.8]{Haenggi2013}. We have illustrated that the \ac{PPP} cannot capture the distribution of inter-vehicle distances, because it permits unrealistically small distances with high probability. The \ac{PPP} becomes more problematic for the left lane because due to the high speeds over there, the drivers maintain large safety distances. We have illustrated that the envelope of the J-function for the fitted hardcore process $\Phi_c$ can capture the J-function of the real snapshot, see~\cite[Sec. IV]{Koufos2019} for a detailed description of the fitting method. 

\begin{figure*}[!t]
 \centering
  \subfloat[Outage probability, $\eta\!=\!3$]{\includegraphics[width=2.25in]{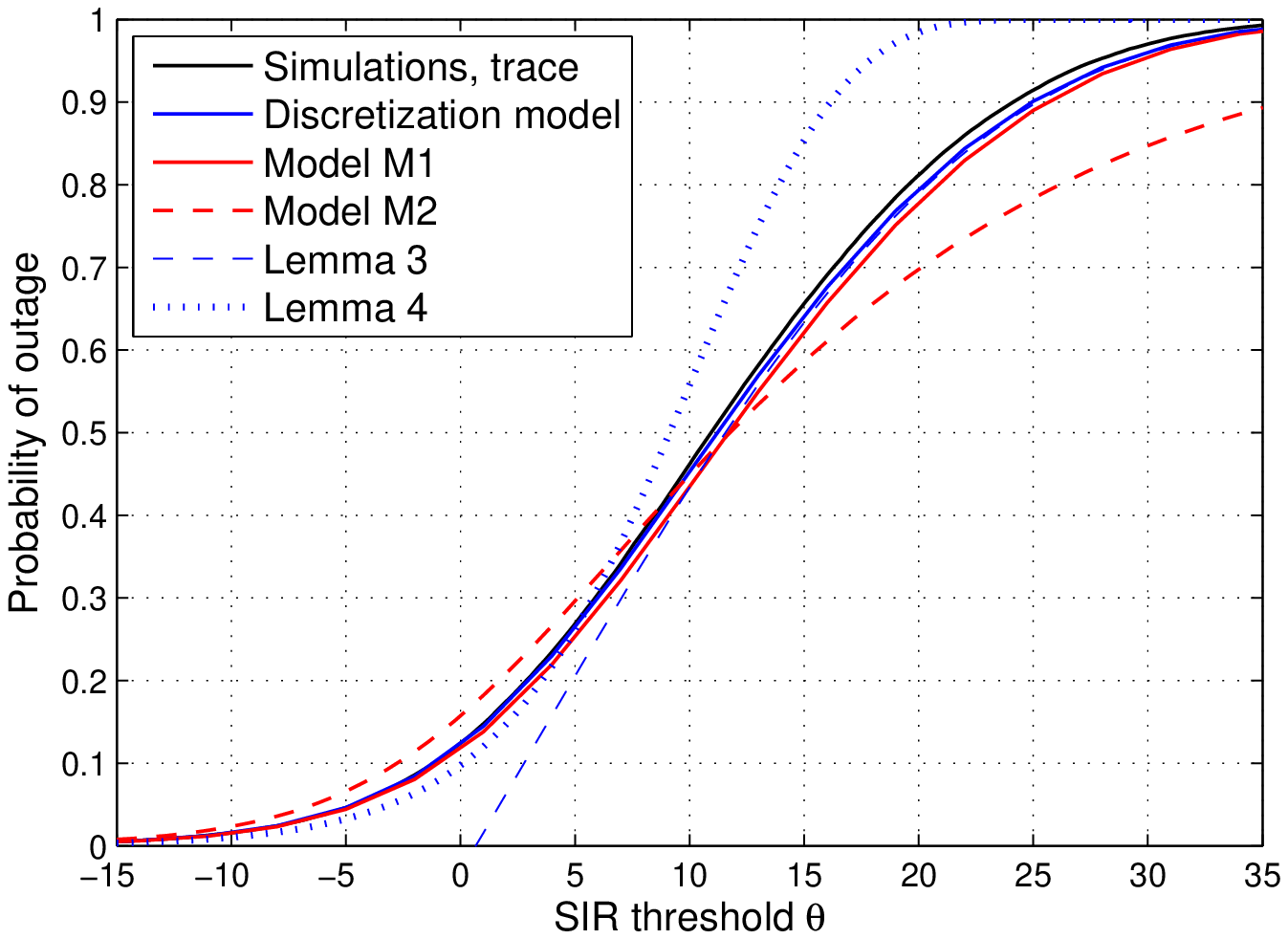}}\hfil
  \subfloat[Meta distribution, $\eta\!=\!3$]{\includegraphics[width=2.25in]{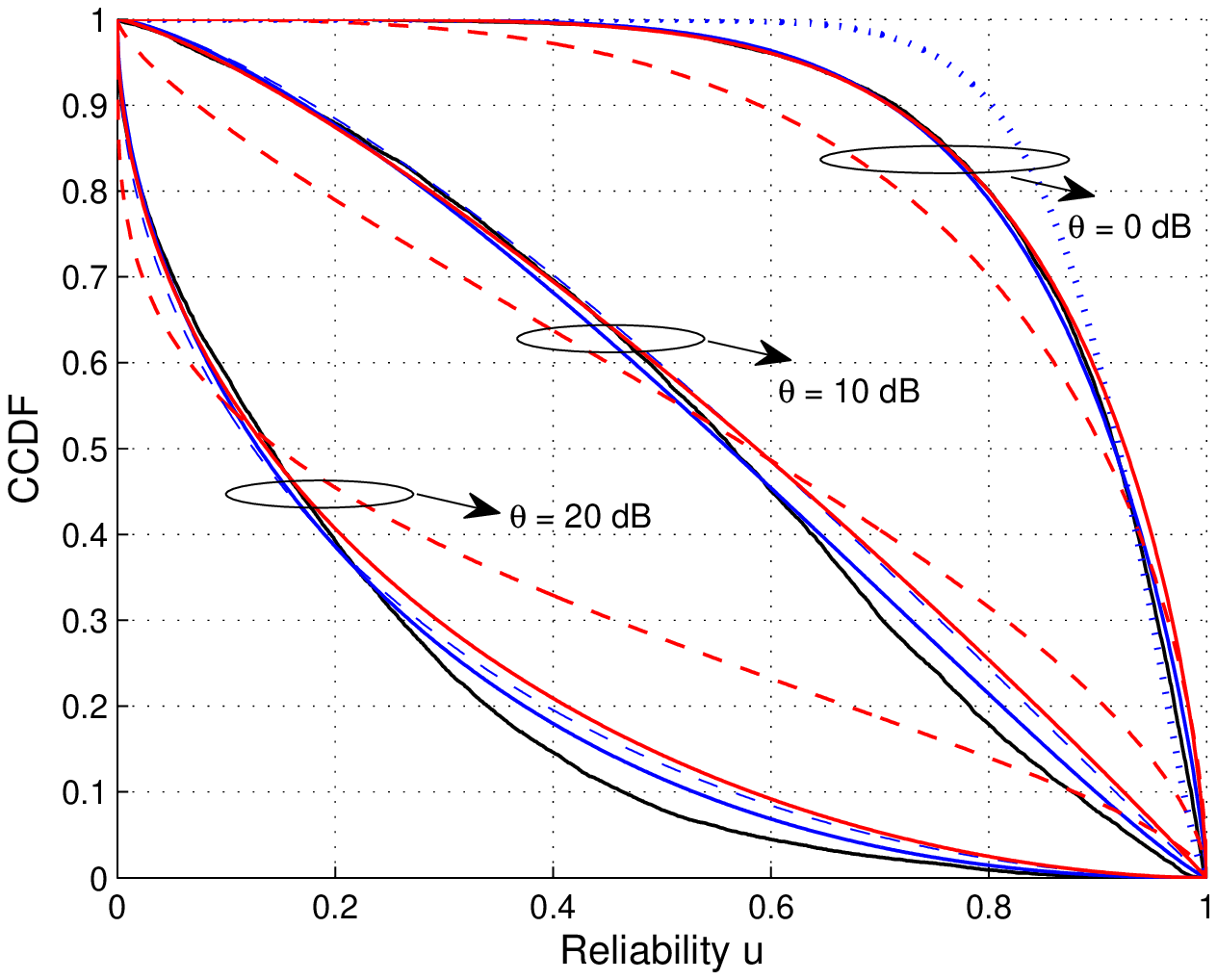}}\hfil
\subfloat[Meta distribution, $\eta\!=\!4$]{\includegraphics[width=2.25in]{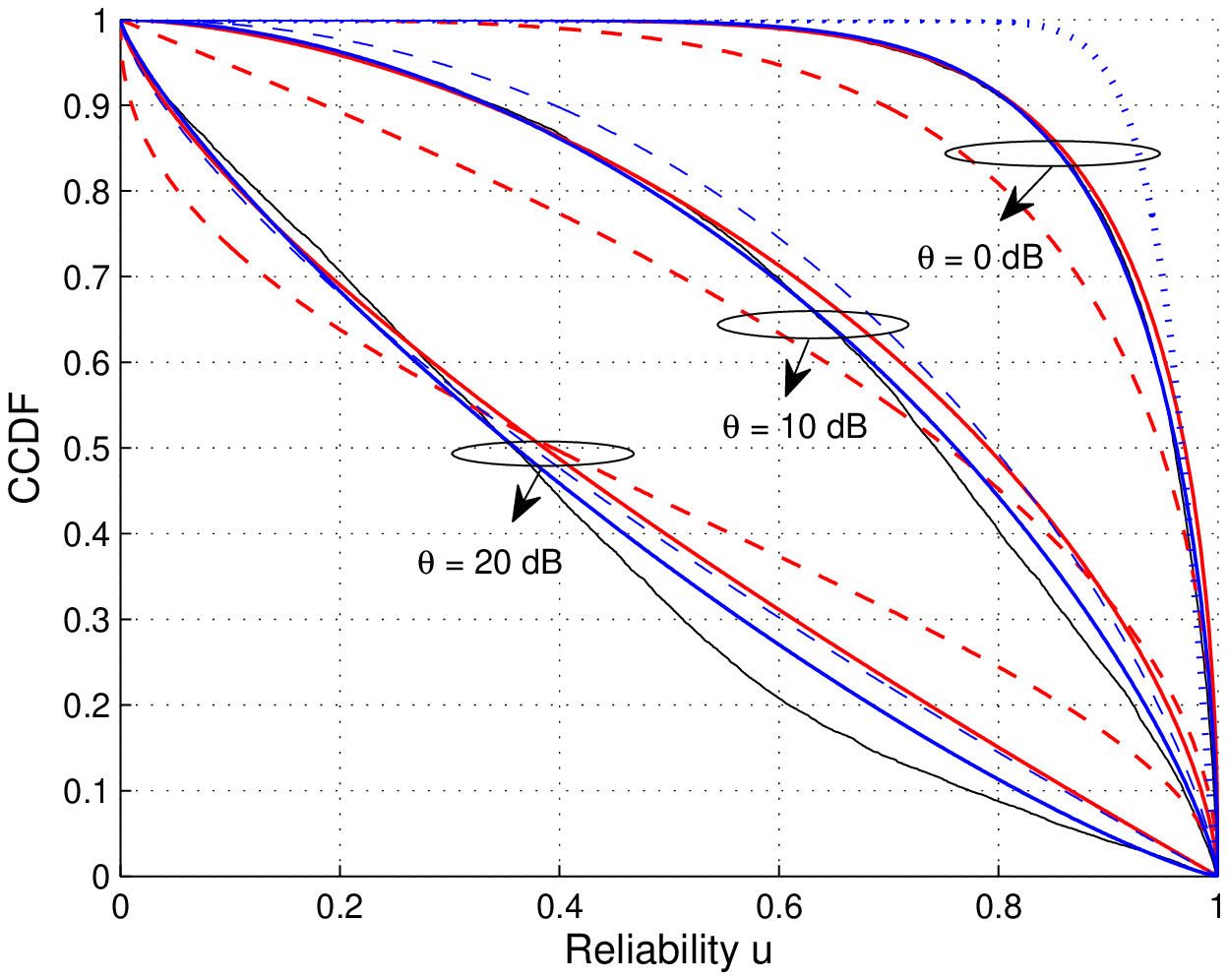}}\hfil
 \caption{(a) The simulated probability of outage using real trace and the approximations using the discretization model, and the models M1 and M2. For the discretization model and M1, the (fitted) shifted-exponential \ac{PDF} has $\lambda\!\approx\!0.0205 \,  {\text{m}}^{-1}$ and $c\!\approx\! 14.82 \, {\text{m}}$, yielding $\lambda c\!\approx\! 0.3037$. For M2, the (fitted) exponential \ac{PDF} has $\lambda\!\approx\!0.0195 \, {\text{m}}^{-1}$. For $q\!=\!2\%$, we get $R_{min}\!\approx\!442 \, \text{m}$ from~\eqref{eq:ThresholdR}, and we calculate $K\!=\!\lceil\frac{R_{min}}{c}\rceil\!=\!30$ and $R\!=\!K c\!\approx\!445 \, {\text{m}}$. $10^5$ simulations. The approximation~\eqref{eq:Lem3M1} in Lemma~\ref{lemma:3} for large thresholds is accurate for $\theta\!>\! 10$,  and the approximation~\eqref{eq:Lem4M1} in Lemma~\ref{lemma:4} for small thresholds is accurate for $\theta\!<\!0.5$. (b)-(c) The simulated meta distribution and the approximations using the same models. $10^4$ spatial configurations and  $10^4$ realizations of fading and activity per configuration. The approximations~\eqref{eq:Lem3M1} and~\eqref{eq:Lem3M2} coupled with the Beta approximation are depicted for large thresholds $\theta\!=\!10$ and $\theta\!=\!100$. For $\theta\!=\!1$, we used the approximations~\eqref{eq:Lem4M1} and~\eqref{eq:Lem4M2}, which are not accurate because Lemma~\ref{lemma:4} assumes small $\theta\!\ll\! 1$. Activity $\xi\!=\!0.5$. We have selected the 1000-th snapshot from motorway M40, May 7 2010, busy hour~\cite{Fiore2014,Fiore2015}. The empirical \ac{CDF} of inter-vehicle distances, obtained from the trace, is used to simulate the locations of interferers and the useful link distance.}
 \label{fig:DataPout}
\end{figure*}

\begin{figure*}[!t]
 \centering
  \subfloat[Outage probability, $\eta\!=\!3$]{\includegraphics[width=2.25in]{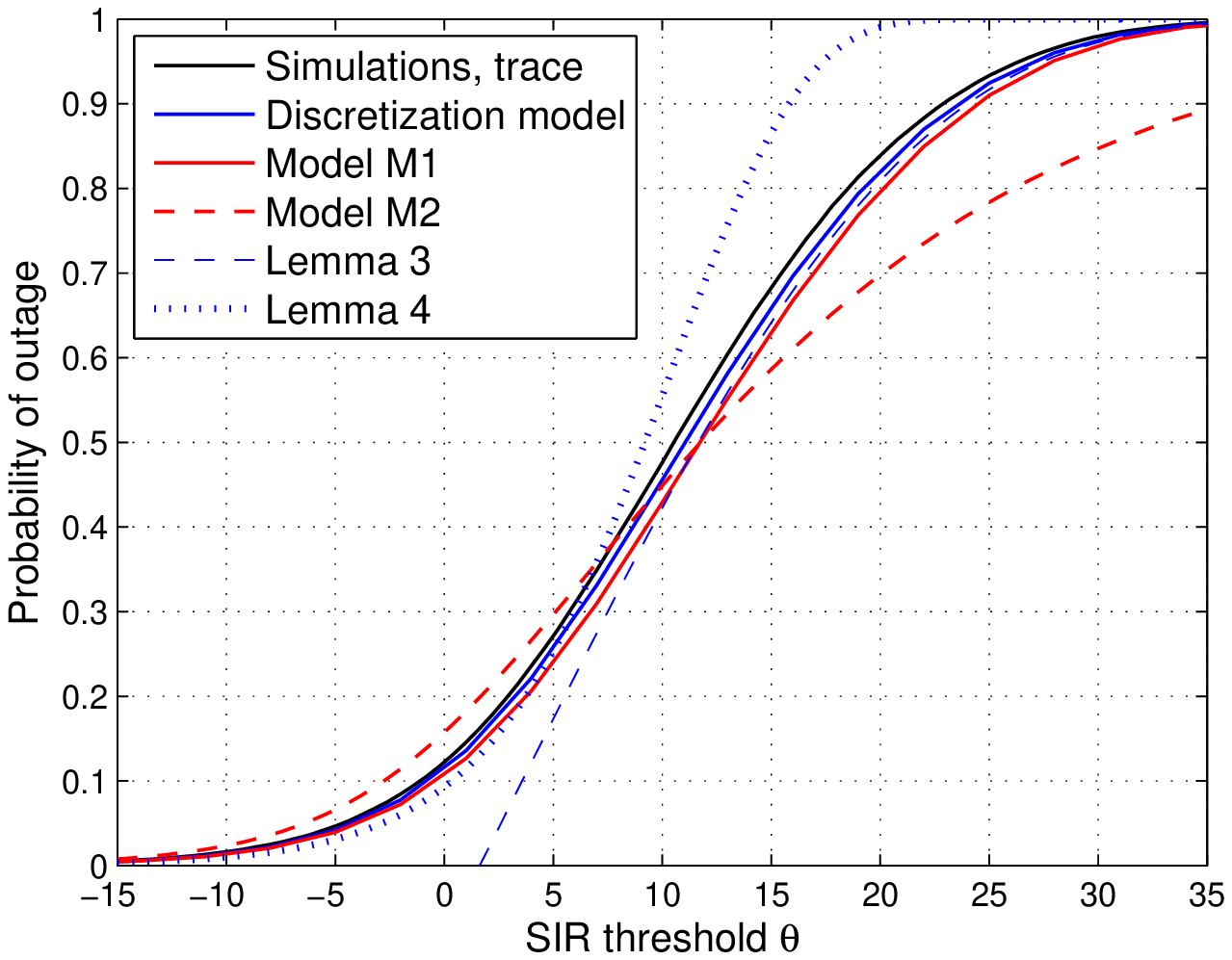}}\hfil
  \subfloat[Meta distribution, $\eta\!=\!3$]{\includegraphics[width=2.25in]{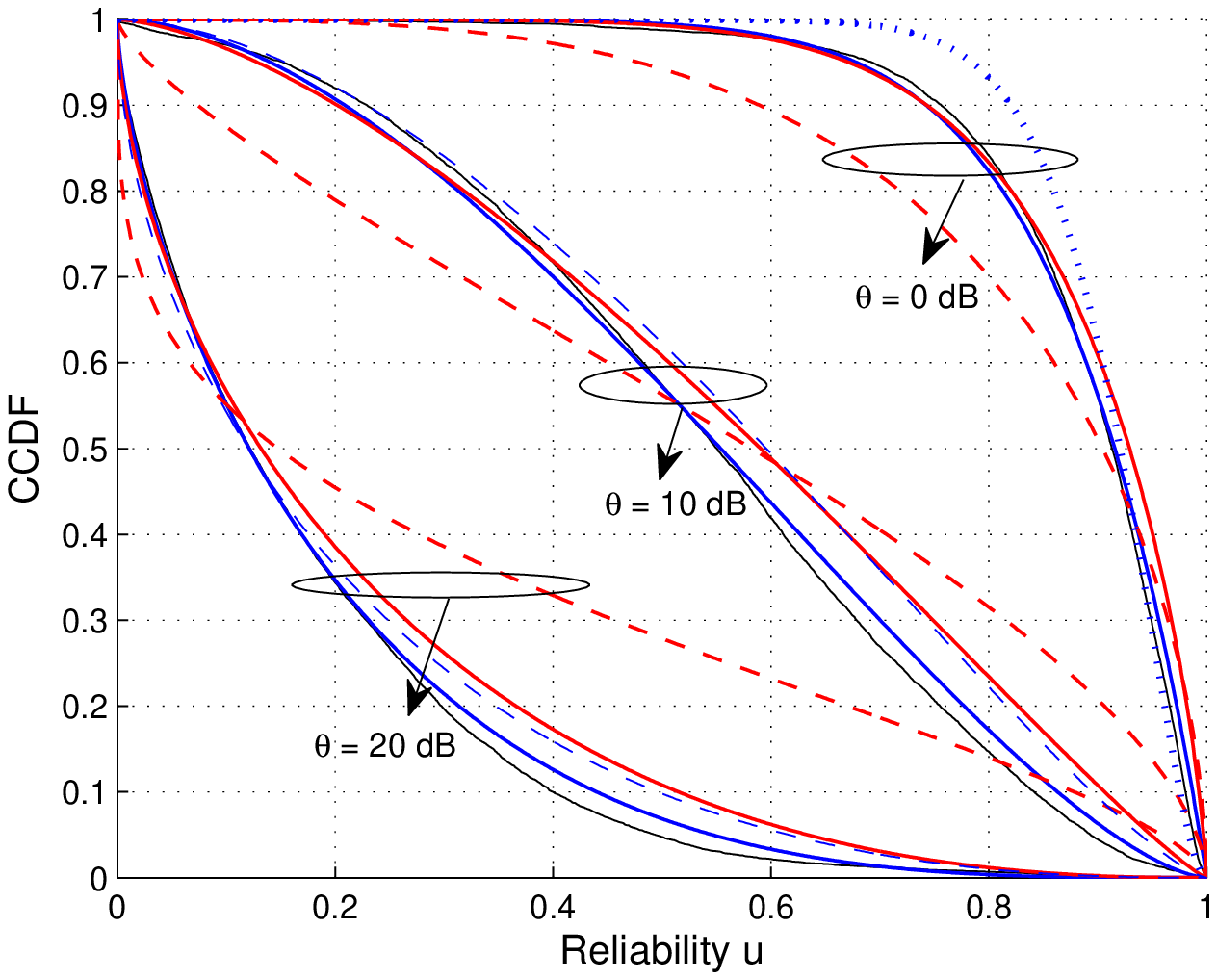}}\hfil
\subfloat[Meta distribution, $\eta\!=\!4$]{\includegraphics[width=2.25in]{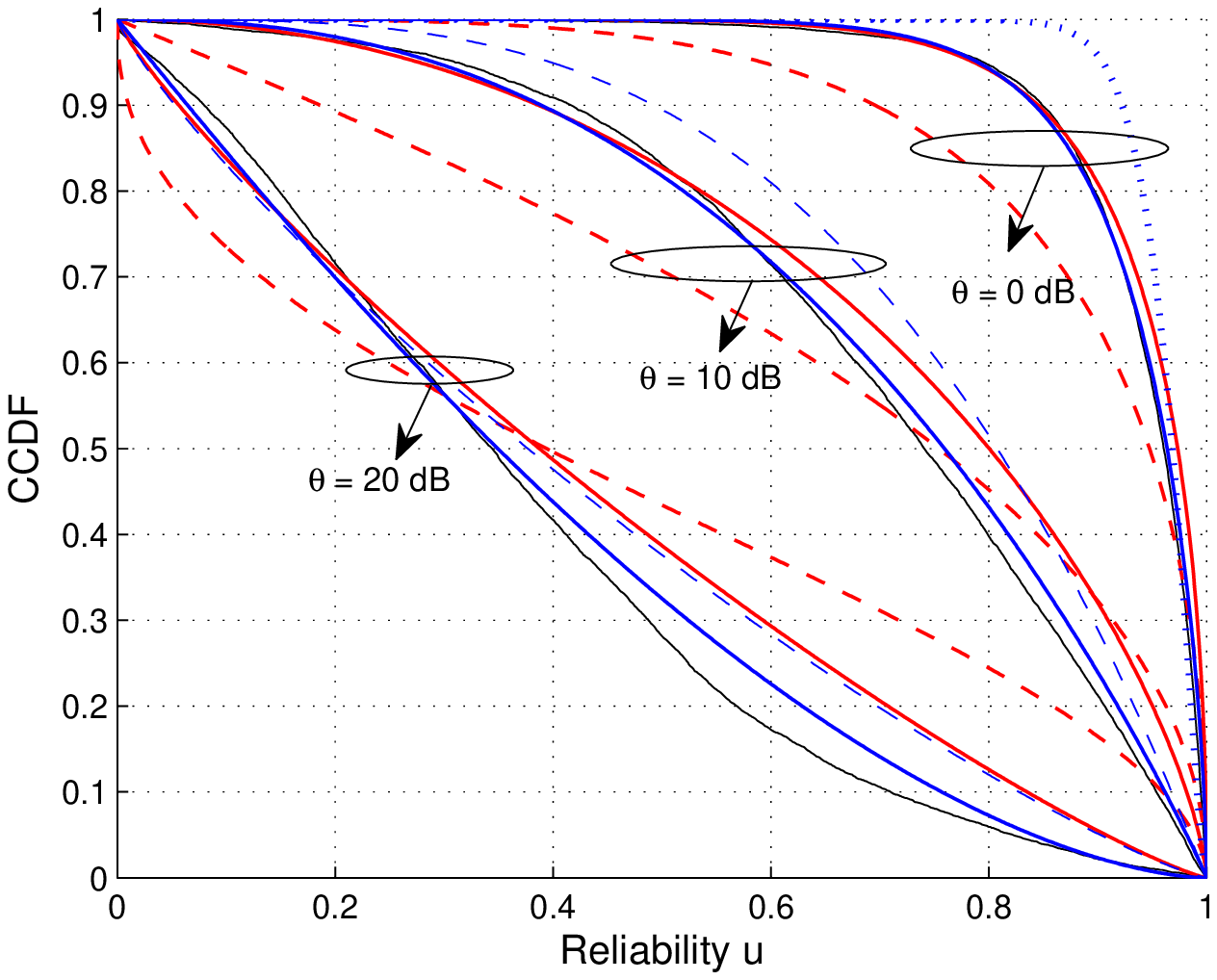}}\hfil
 \caption{Fitting the models to the simulations using the 1200-th snapshot from motorway M40, May 7 2010, busy hour~\cite{Fiore2014,Fiore2015}. For the discretization model and the model M1 we use the estimates $\lambda\!\approx\!0.0203 \,  {\text{m}}^{-1}, c\!\approx\!19.76 \, {\text{m}}$, resulting to $\lambda c\!\approx\! 0.4017$. For the model M2 we estimate $\lambda\!\approx\!0.0191 \, {\text{m}}^{-1}$. See the caption of Fig.~\ref{fig:DataPout} for explanation of the legend and other parameter settings.}
 \label{fig:DataPout1200}
\end{figure*}

In the current paper, we use the fitted parameters $\lambda, c$ to assess which model (discretization, M1 and M2) can better predict the simulated outage probability and the meta distribution of snapshots. We see in Fig.~\ref{fig:DataPout} that for the left lane of the selected snapshot, $\lambda c\!\approx\! 0.3$, the discretization model outperforms the model M1, especially for large thresholds, while M2 incurs very large errors. Higher spatial regularity, $\lambda c\!\approx\! 0.4$ in Fig.~\ref{fig:DataPout1200}, makes clear the benefit of discretization. 

\begin{figure}[!t]
 \centering
  \includegraphics[width=3in]{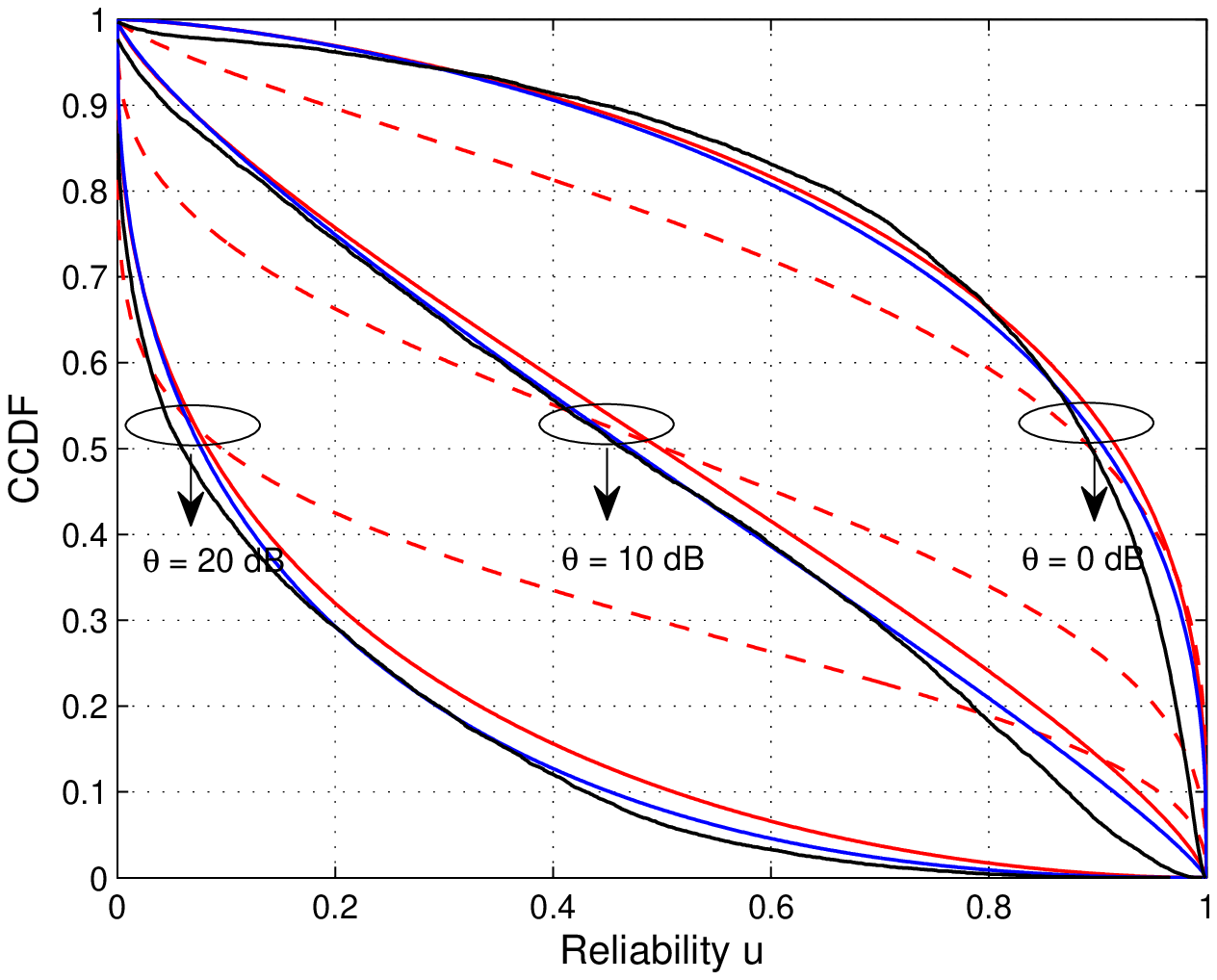}
 \caption{Meta distributions for the 1200-th snapshot from motorway M40, May 7 2010, busy hour, left lane, considering interference from all three lanes. For lane separation $4 \, {\text{m}}$ and antenna beamwidth $\frac{\pi}{20}$ we have $r_0\!\approx\! 50 \, {\text{m}}$ for the middle lane and $2r_0\!\approx\! 100 \, {\text{m}}$ for the right lane. For the right lane, we assume a PPP beyond $2r_0$ for all three models with $\lambda_r\!\approx\! 0.0205 \, {\text{m}}^{-1}$. For the middle lane we assume \ac{PPP} beyond $r_0$ for the models M1 and M2 with $\lambda_m\!\approx\! 0.0186 \, {\text{m}}^{-1}$. For the discretization model we estimate $\lambda_m\!\approx\! 0.0193\,  {\text{m}}^{-1}, c_m\!\approx\!9.86 \, {\text{m}}$, and $K_m\!=\!14$ for $q\!=\! 2\%$. $\eta\!=\!4$, see the caption of Fig.~\ref{fig:DataPout1200} for other parameter settings and mapping of line styles to models.}
 \label{fig:DataMultiLane}
\end{figure}

The extension of the discretization model to multiple lanes is straightforward by discretizing with lane-specific parameter $c$. Due to directionality, only vehicles behind some distance $r_0$ from the receiver may contribute to other-lane interference, see~\cite[Section VI]{Koufos2019}. The interference originated from other lanes does not require to constrain the location of any of the points of the processes. We just add a reference point (the receiver) at the origin. The separation threshold between near- and far-field interferers for other lanes is easier to calculate. Starting from $\int_R^\infty r^{-\eta}{\rm d}r\!\leq\!q\int_{r_0}^\infty r^{-\eta}{\rm d}r$, we have $R\!\geq\!\left(1/q\right)^{\frac{1}{\eta-1}} r_0$. The results are available in Fig.~\ref{fig:DataMultiLane}. Summing up, Fig.~\ref{fig:DataPout} $-$ Fig.~\ref{fig:DataMultiLane} highlight the discretization model as a reasonable choice for modeling the outage probability along motorway \acp{VANET}. 

\section{Conclusions}
\label{sec:Conclusions}
In high-speed motorways the vehicles maintain large safety distances from the vehicle ahead. Because of that, a shifted-exponential \ac{PDF}  captures the distribution of headway distance along a lane much better than the \ac{PPP}. In order to approximate the \ac{PGFL} of the shifted-exponential (or hardcore) process, we used a shifted-exponential \ac{PDF} for the  link distance coupled with a guard zone (equal to the hardcore distance) behind the transmitter. This model predicts the moments of the \ac{SIR} much better than the \ac{PPP}. Nevertheless, for regular deployments, and/or large \ac{SIR} threshold and transmission probability, it starts to lose some of its power. Because of that, we have devised a discretized deployment for the near-field. This is more tailored to the hardcore constraints, hence it approximates better the \ac{PGFL} of the hardcore process. The discretization model coupled with the Beta \ac{PDF} for the meta distribution can capture the statistics of the outage probability along a lane of the motorway. This has been validated against synthetic traces generated from real vehicular data, considering also interference due to multiple lanes. The discretization model would be useful for network planning, because it provides the system designer with better prediction of the distribution of outage probabilities than that obtained with a \ac{PPP}.

We have shown that for increasing \ac{SIR} thresholds the disparity in the success probability across different links increases. This means that the upper tail of the \ac{SIR} \ac{CDF}, calculated as an average over all spatial realizations, does not represent accurately the outage probability along a snapshot of the motorway. As a result, the calculation of the meta distribution becomes useful. We have also shown that the \ac{PPP} predicts much higher disparity than that observed with real traces, because it allows more variability in the link distance. Another important conclusion is that the mean and the variance of the meta distribution in the low \ac{SIR} depend only on the product of transmission probability and \ac{SIR} threshold. Therefore lowering the activity by $x$ dB allows increasing the operation threshold by $x$ dB with little effect on the meta distribution. This is, however, not the case for high \ac{SIR} thresholds. Regarding future work, it would be good to see the calculation of the meta distribution also for the rate statistics. In addition, it would be interesting to study the joint distribution of the \ac{SIR} over multiple slots, which is related to the design of retransmission schemes for \acp{VANET}. Some preliminary analysis about the temporal statistics of interference under the hardcore point process can already be found in~\cite{Koufos2019c}.

\section*{Acknowledgment}
The authors would like to thank Harpreet Dhillon for helpful discussions.

%% \begin{IEEEbiography}{Konstantinos Koufos}  obtained the diploma in electrical and computer engineering from Aristotle University, Greece, in 2003, and the M.Sc. and D.Sc. in radio communications from Aalto University, Finland, in 2007 and 2013. After working as a post-doctoral researcher in Aalto University, he has been with the School of Mathematics in the University of Bristol, Bristol UK, as a senior research associate in spatially embedded networks. His current research interests include stochastic geometry, interference and mobility modeling for 5G  wireless networks. 
%% \end{IEEEbiography}
%% \begin{IEEEbiography}{Carl P. Dettmann}
%% received the BSc (Hons) and the PhD in physics from the University of Melbourne, Australia, in 1991 and 1995, respectively. Following research positions at New South Wales, Northwestern, Copenhagen and Rockefeller Universities he moved to the University of Bristol, Bristol, UK, where he is now Professor and Deputy Director of the Institute of Probability, Analysis and Dynamics.  Professor Dettmann has published over 120 international journal and conference papers in complex and communications networks, dynamical systems and statistical physics. He is a Fellow of the Institute of Physics and serves on its fellowship panel.  Professor Dettmann has delivered many presentations at international conferences, including a plenary lecture at Dynamics Days Europe and a Tutorial at the International Symposium on Wireless Communication Systems.
%% \end{IEEEbiography}
\end{document}